\documentclass[a4paper,UKenglish,cleveref, autoref]{lipics-v2019}
\usepackage{hyperref}
\usepackage{amssymb}
\usepackage{amsmath}
\usepackage{latexsym}
\usepackage{multicol}
\usepackage{booktabs}
\usepackage{multirow}
\usepackage{comment}
\usepackage{amsthm}
\usepackage[ruled,linesnumbered]{algorithm2e}

\title{Continuous Histograms for Anisotropy of 2D Symmetric Piece-wise Linear Tensor Fields} %TODO Please add

\titlerunning{Continuous Histograms for Anisotropy}
\author{Talha Bin Masood}{Link\"oping University, Sweden}{talha.bin.masood@liu.se}{https://orcid.org/0000-0001-5352-1086}{}
\author{Ingrid Hotz}{Link\"oping University, Sweden}{ingrid.hotz@liu.se}{https://orcid.org/0000-0001-7285-0483}{}

\authorrunning{T. B. Masood and I. Hotz}
\Copyright{Talha Bin Masood and Ingrid Hotz}
\nolinenumbers %uncomment to disable line numbering

\hideLIPIcs  %uncomment to remove references to LIPIcs series (logo, DOI, ...), e.g. when preparing a pre-final version to be uploaded to arXiv or another public repository

\providecommand*{\anis}{\ensuremath{\alpha}}
\providecommand*{\sqanis}{\ensuremath{\nu}}

\begin{document}

%\titlerunning{Continuous Histograms for Anisotropy}
%\author{Talha Bin Masood and Ingrid Hotz}
%\institute{Talha Bin Masood, Ingrid Hotz \at Department of Science and Technology (ITN), Link\"oping University, Sweden.\\ \email{\{talha.bin.masood, ingrid.hotz\}@liu.se}}
%\and Name of Second Author \at Name, Address of Institute \email{name@email.address}}
\maketitle

\abstract{
 The analysis of contours of scalar fields plays an important role in visualization. For example the contour tree and contour statistics can be used as a means for interaction and filtering or as signatures. In the context of tensor field analysis, such methods are also interesting for the analysis of derived scalar invariants. While there are standard algorithms to compute and analyze contours, they are not directly applicable to tensor invariants when using component-wise tensor interpolation. In this chapter we  present an accurate derivation of the contour spectrum for invariants with quadratic behavior computed from two-dimensional piece-wise linear tensor fields. For this work, we are mostly motivated by a consistent treatment of the anisotropy field, which plays an important role as stability measure for tensor field topology. We show that it is possible to derive an analytical expression for the distribution of the invariant values in this setting, which is exemplary given for the anisotropy in all details. Our derivation is based on a topological sub-division of the mesh in triangles that exhibit a monotonic behavior. This triangulation can also directly be used to compute the accurate contour tree with standard algorithms. We compare the results to a na\"ive approach based on linear interpolation on the original mesh or the subdivision.
}

%% SECTION 1 Introduction
%% ========================================================================= %%
\vspace{3em}
\section{Introduction} \label{section:introduction}

Contours or isosurfaces of scalar fields play a central role in visualization. We will only use the term contour in this chapter. There is a large body of work centered around contour computation and analysis with many applications. A prominent example is the {\em contour tree}, which provides a structural overview of the data. Complementary information is provided by some kind of {\em contour statistics}, sometimes also referred to as {\em continuous histogram}, which is an extension of the discrete histogram to encode the distribution of the scalar values of continuous functions. It can serve as a valuable quantitative signature~\cite{Bajaj1997} of a data set. Both structures are frequently used for interaction and filtering, for example, to determine interesting iso-values or for transfer function design~\cite{Weber2007}. Histograms have also been applied for tensor field visualization and exploration to display the distribution of scalar invariants derived from tensor fields~\cite{Kratz2011}. The {\em join tree}, a subset of the contour tree, of the anisotropy builds the foundation for a stability measure of topological features in the tensor field~\cite{Wang2017}. There are many standard algorithms to compute and analyze contours mostly based on piece-wise linear or multi-linear behavior. This assumption is, however, often violated for tensor invariants when using component-wise tensor interpolation. In this chapter, we present the derivation of an exact closed-form expression for {\em{continuous histograms}} for quadratic tensor invariants that is consistent with a piece-wise linear interpolation of a tensor filed given over a two-dimensional triangulated domain. Thereby we are especially interested in the anisotropy, an important characteristic in many applications, defined as the difference between major and minor eigenvalue. 

Our main contributions are:
\begin{itemize}
    \item Framework for the analysis of scalar invariants that is consistent with a piece-wise linear interpolation of the tensor components. This includes topologically correct contours, the correct contour tree, and the exact contour histograms.
    \item A generic closed-form formulation of histograms for continuous data based on a generalization of the cumulative histogram. This approach neither requires explicit contour length computations nor the computation of the gradient. 
    \item Explicit solution for the histogram for piece-wise quadratic functions with positive Hessian over a two-dimensional domain. The anisotropy over a piece-wise linear component interpolation provides a relevant example.
    \item Comparison to na\"ive approaches using a linear interpolation of the anisotropy and a discussion of the results.
\end{itemize}

In the remainder of this section, we first summarize the development of the concept of histograms from a signature of discrete data to continuous fields. Then we summarize some aspects of tensor interpolation and invariants, that motivated our work.

\subsection{Continuous histograms and contour statistics}
Originally \emph{histograms} have been developed for discrete data. They go back to Pearson and refer to a graphical representation displaying the frequency of the data items as bars over the data range~\cite{Ioannidis2003}. As such a histogram provides a summary of the data and they are also used to compare and characterize data sets. 
Formally the histogram for a discrete set of data values $F=\{f_i | i \in {1 , \cdots , m}\}$
is a one-dimensional bar chart displaying the frequency of each distinct element $f\in F$ by a bar with height \[h(f)= \sum_i\delta(f-f_i) \]
where $\delta$ is the delta function which is equal to one if $f=f_i$ and zero otherwise. 
A related concept is the \emph{cumulative histogram}, applicable to ordered data, counts the cumulative number of data values smaller than a given value $f\in F$. This results in values 
\[c(f)= \sum_{f_i\le f}h(f_i).\]  
Today, the term is often not only used for the graphical representation but also for the concept capturing the distribution of function values by binning the function range and counting the frequency of data samples within these bins. 
The resulting histograms, however, are strongly dependent on the binning size and the sampling strategy of the function in the domain. The continuous nature of the function between the sample points is not represented. For these reasons several concepts to extend histograms to continuous functions have been proposed resulting in a distribution of the function values. 
Bajaj{\em ~et~al.}~\cite{Bajaj1997} have introduced the \emph{contour spectrum} as a data-signature to find interesting iso-values for volume rendering. The contour spectrum assigns the geometric measures of the contour length, respectively surface area, to each scalar value. In addition, they consider areas and volumes of regions below or above a given iso-value. They also propose a method to compute these values exactly under the assumption of a piece-wise linear interpolation and a piece-wise constant gradient.
Later Carr{\em ~et~al.}~\cite{Carr2007} investigated the relation between histograms and the contour spectrum based on a nearest-neighbor interpolation. Scheidegger{\em ~et~al.}~\cite{Scheidegger2008} completed this work and introduced a natural generalization of histograms to the continuous setting as the distribution given by the contour spectrum weighted by the ``local isosurface density'' expressed by the inverse gradient magnitude. Given a scalar field $f$ and scalar value $h$ the distribution is given as
\[\pi_f(h) = \int_{f^{-1}(h)}|\nabla f(x)|^{-1}dS\]
For the derivation of the distributions, they refer to Coarea formula that provides a relationship between the sum of area integrals and a global volume integral as used by Mullen{\em ~et~al.}~\cite{Mullen2007}. For the computation of the distribution, they propose an approximation based on the count of active cells weighted by the inverse of the cell span, the difference between max and min per cell, to approximate the gradient. For piece-wise linear function, this yields a more or less good approximation of the real distribution.

Similar concepts have also be considered for multivariate fields, for example, generalizations of scatterplotts~\cite{Bachthaler2008} and parallel coordinates~\cite{Heinrich2009} to continuous data. 
In both cases, the mathematical model is based on mass conversation of a mapping from the function range to the scatterplot space, respective parallel coordinates space. The continuous histogram is a special case considering a one-dimensional range. 
While the mathematical model is generic, the proposed computational algorithms are limited to specific interpolation models in the spatial domain. 
The algorithm by Bachthaler{\em ~et~al.}~\cite{Bachthaler2008} assumes a linear interpolation in tetrahedral cells.
The work by Heinrich{\em ~et~al.}~\cite{Heinrich2009} introduces a splatting approach to generate the distributions by computing footprints for Gaussian input kernels.
These limitations have also been discussed by Bachthaler{\em ~et~al.}~\cite{Bachthaler2009}. They propose an adaptive refinement approach that does not require a linear interpolation.  
The algorithm, however, still assumes that the interpolation is convex which is not the case for most tensor invariants based on a component-wise tensor interpolation.

In our work, we focus on invariants with a quadratic behavior and propose an exact solution pursuing a slightly different approach than most of the previous methods.
Instead of directly generalizing the concept of histograms we start with the generalization of the cumulative histogram. 
This approach requires neither surface area or contour length nor gradient approximations and is thus much more flexible in terms of requirements for the input functions and interpolation schemes.
The continuous scatterplot or contour distribution follows than as a derivative of the cumulative histogram. 

\subsection{Notes on tensor field interpolation}
In this section, we briefly summarize some notes related to tensor field interpolation without going much into detail. This has been a frequently discussed topic in many applications and it is agreed that this is a challenging topic. 

While many theories and visualization models assume data given as fields on continuous domains, the data-reality are data sets sampled at discrete locations. Depending on the origin of the data this can be voxel-based data from imaging or meshes for simulation data. In any case, it is necessary to approximate and reconstruct the field from these samples.
Thereby the most commonly used methods for tensor fields are based on a component-wise interpolation. Similar to other applications and other data types these are linear, bi-, or tri-linear interpolation depending on the grid type. 
Concerning such interpolations, a frequently discussed topic is the non-linear dependency of most tensor invariants on the tensor components~\cite{Kindlmann2007}. There have been many methods proposed which explicitly try to preserve certain tensor characteristics~\cite{Batchelor2005, Sreevalsan-Nair2011}. Recently there appeared a survey over interpolation methods for positive definite tensors~\cite{Feragen2017}.

We will, however, stay with the most simple interpolation methods, linear component-wise interpolation within tetrahedra, a method that is also often the basis for finite element simulations.
The non-linear dependence of the tensor invariants can lead to a non-convex behavior of the invariants inside the cells. 
An example, we are especially interested in, is the behavior of the anisotropy, defined as the difference between the maximum and minimum eigenvalue, which is typically decreased inside the cells. 
Sometimes this is referred to as ``swelling-artifact'', but we rather argue that this is a fundamental property of tensor behavior. It reflects the fact that the anisotropy is linked to the directional behavior of the tensor. This is expressed by the fact that the anisotropy is always zero at degenerate points and can directly be used to measure the stability of the degenerate points~\cite{Jankowai2019}. Therefore the analysis of the anisotropy field must be handled consistently with the chosen interpolation method for the tensor field. This concerns the computation of contours, the contour tree and also histograms.

For our consideration in this work, we assume that we have a continuous tensor field given on a two-dimensional triangulated domain. We further assume a piece-wise linear component-wise linear interpolation of the tensor field within these triangles. The anisotropy has then a quadratic behavior in the triangle. One can observe similar behavior for other invariants, for example, the determinant too. All the above-proposed methods for histogram computations fail when the interpolation of the data is not convex. Similarly, typical contour or contour tree computation is also based on convex interpolation. 

\subsection{Contour trees, a topological summary of scalar functions}
\begin{figure}[ht]
\centering
\includegraphics[height=4cm]{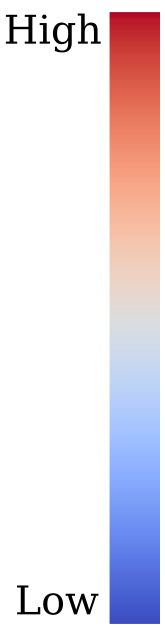}
\includegraphics[height=4.1cm]{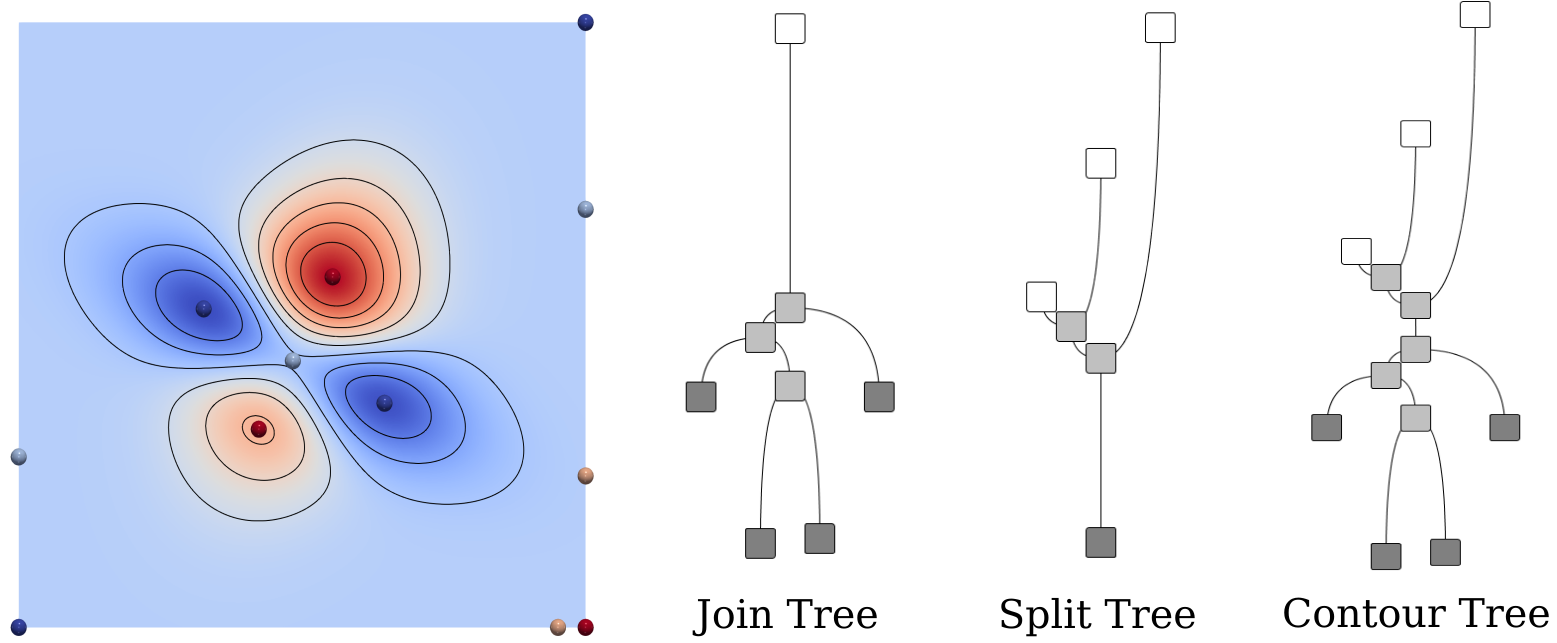}
\caption{For the scalar field on the left, the join tree, split tree and the contour tree are shown.}
\label{fig:contourtree}
\end{figure}
The contour tree keeps track of the changes of sub- and superlevel sets of a function and thus provides a valuable summary of the function. It can be assembled form the {\em join tree} tracing the changes in contours when the function value is increased from $-\infty$ to $\infty$ and the {\em split tree} one where the function value is decreased from $\infty$ to $-\infty$. In more detail, the join tree tracks the creation and merging of sublevel sets, which are recorded in a tree. At the local minima of the function, new branches are created. As the function value increases, branches are extended and merged at saddle points where two sublevel sets merge. The global maximum of the function becomes the root of the tree. The split tree is constructed accordingly. The most commonly used algorithm for contour tree computation assumes a piecewise linear interpolation of the scalar field where all critical points are located at the cell vertices~\cite{Carr2003}. 
In the case of the anisotropy, we are especially interested in the join tree since many of the minima correspond to the degenerate points in the tensor field. The join tree can be used to quantify the stability of these points~\cite{Wang2017}. For correct results an accurate join tree computation based on the quadratic behavior or the anisotropy is essential.
%% SECTION 3 Problem
%% ========================================================================= %%
\section{Problem statement and Solution Overview} \label{section:overview}
Given a tensor field sampled over a 2D triangular mesh, our goal is to compute the accurate contour-tree and histogram of the tensor anisotropy consistent with linear interpolation of tensor components. Anisotropy is a quadratic function under this interpolation. 

\begin{figure}[ht]
 \centering
 \includegraphics[width=0.7\linewidth]{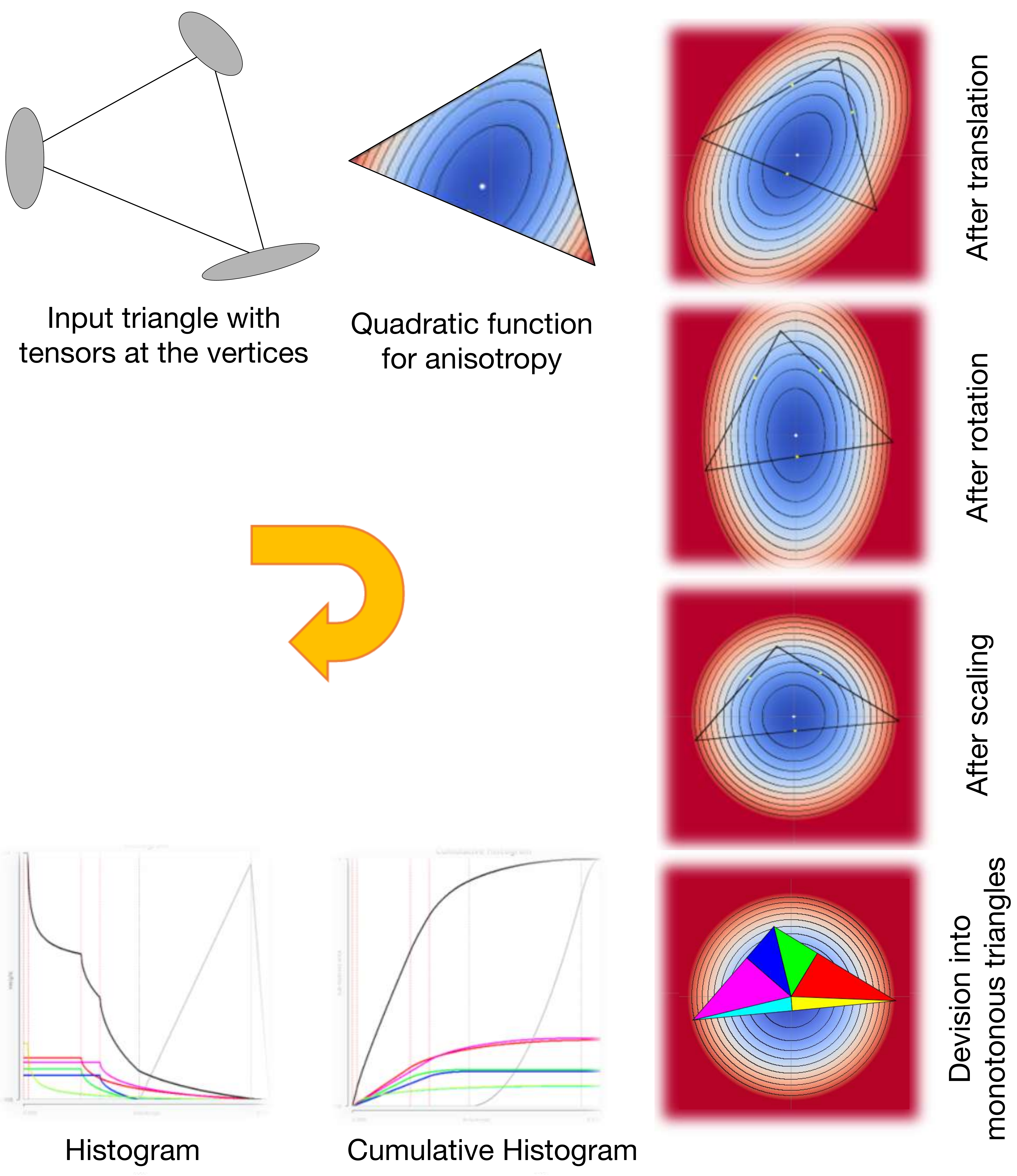}
  \caption{Solution overview: First, the coefficients of the quadratic function are determined. Then a transformation is applied such that the minimum is at the origin and the contours are circular, followed by  a sub-division into monotonous triangles.
 For each monotonous triangle, the cumulative anisotropy histogram is computed using sub-level set areas, which are added to get the cumulative histogram for the original triangle. Finally, the derivative of the cumulative histogram yields the anisotropy histogram.  
  }
  \label{fig:overview}
\end{figure}

Figure~\ref{fig:overview} gives an overview of our solution approach. For each triangle in the input mesh, we first determine the coefficients of the quadratic function for the anisotropy based on tensor values at the triangle vertices. We show that anisotropy is a special quadratic function which either has a single minimum equal to zero and elliptical contours, or in a degenerate case, minima along a line and the contours are pairs of parallel lines, refer to Section~\ref{section:anisotropy}. The first step is the subdivision of the triangle into monotonous sub-triangles, which is the basis for the topologically correct contour extraction, the correct contour tree computation and the derivation of the histogram.
We obtain the histogram as the derivative of the cumulative histogram, which is computed first from the area of the sub-level set for the isovalues. To compute the explicit area of the sub-level sets we apply a linear transformation to the coordinate system such that elliptical contours become circular contours centered at the origin, Section~\ref{sec:transforms}.
%The individual steps of the pipeline are discussed in more detail in the following.

%% SECTION 3 Background
%% ========================================================================= %%
\section{Background and Notations} \label{section:background}
In this section, we provide the notations and a brief mathematical background required for the proposed method discussed in sections later. First, in section~\ref{section:back_tensors} we set up the notations for tensors and the invariant of interest to us that is tensor anisotropy. Then in section~\ref{section:back_barycenter}, we describe the barycentric interpolation within a triangle, since this approach is used for linear interpolation of tensors within a triangle. Lastly, in section~\ref{section:back_quadratic}, we discuss the general bi-variate quadratic function, its critical point, and the shape of its contours. 

\subsection{Second order symmetric tensors and anisotropy} \label{section:back_tensors}
For a second order symmetric tensor, 
$
T=
  \begin{pmatrix}
    e & f \\
    f & g
  \end{pmatrix}    
$,
 the two eigenvalues are:
\begin{equation}
    \lambda, \mu = \frac{1}{2}(e+g)\pm\sqrt{\frac{(e+g)^2}{4}-(eg-f^2)}  \nonumber
\end{equation}
Depending on the application different measures for anisotropy are used. For positive definite tensors, relative measures like the fractional anisotropy are common. In mechanical engineering a typical measure is the difference between the maximum and the minimum eigenvalues $\anis(T)= \lambda - \mu = \sqrt{(e+g)^2-4(eg-f^2)}$. It has been shown that this value is also related to the stability of degenerate points in tensor field topology~\cite{Jankowai2019} and is the measure we are mostly interested in.
In the following, we will however consider the squared value of $\anis(T)$, which has the same topological characteristics but simplifies the computations a lot. 

\begin{align}
    \sqanis(T) &= \anis^2(T) = (\lambda - \mu)^2 \nonumber \\ 
             &= (e-g)^2+4f^2  \label{eq:anisotropyGeneral}
\end{align}

\subsection{Barycentric coordinates and piece-wise linear interpolation} \label{section:back_barycenter}
For all our considerations in this paper we use a component wise linear interpolation of the tensor field.
We use barycentric coordinates for interpolation in the triangle.
Since we require an explicit representation of the the tensor field for the derivation of the exact histogram we briefly review the main definitions and formulas in this section. 

\begin{figure}[ht]
\centering
\includegraphics[width=0.5\linewidth]{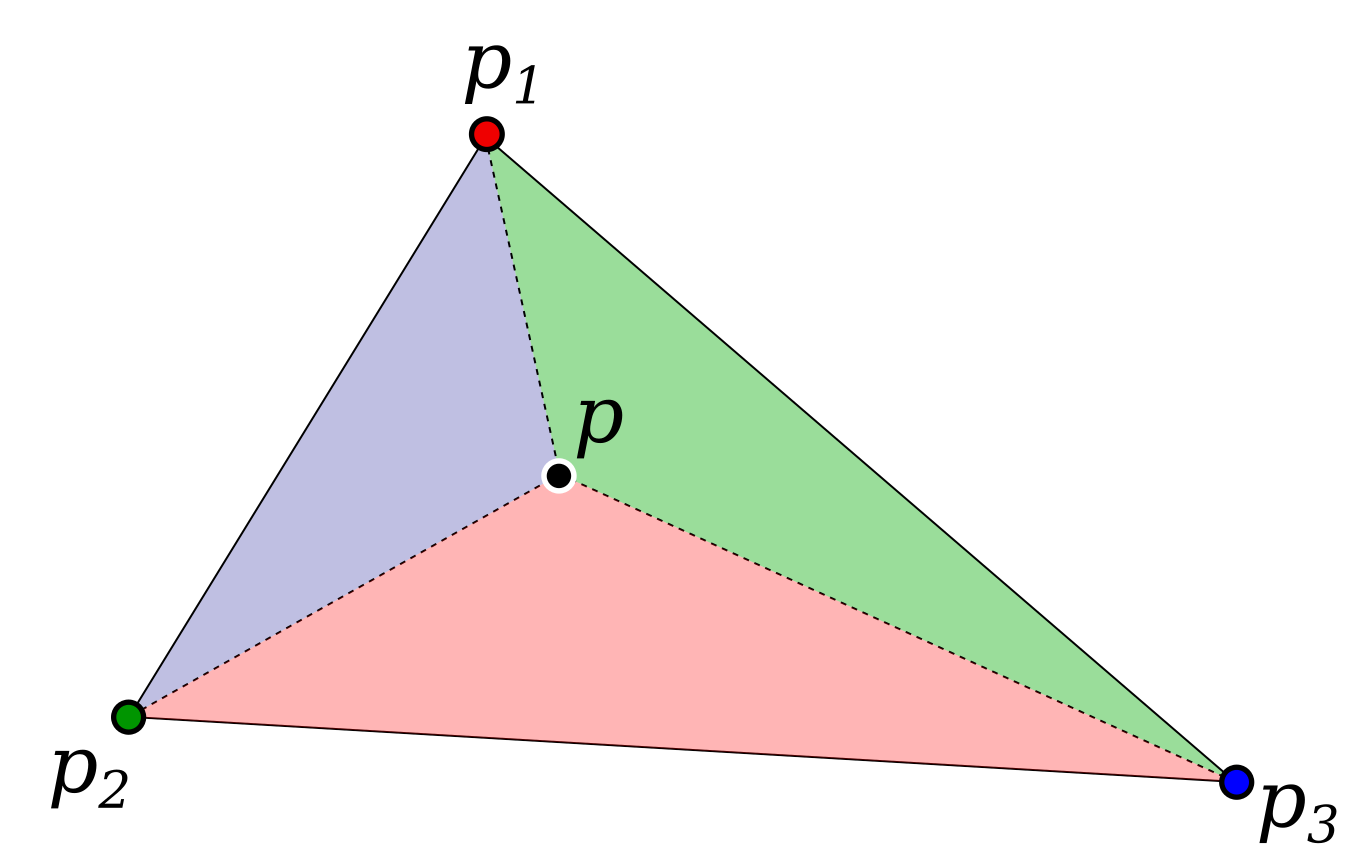}
\caption{Barycentric coordinates and interpolation. The point $p$ within the triangle $\Delta p_1p_2p_3$ can be represented using barycentric coordinates $(\beta_1, \beta_2, \beta_3)$, where $\beta_1=\text{Area}(\triangle pp_2p_3)/\text{Area}(\triangle p_1p_2p_3)$, $\beta_2=\text{Area}(\triangle pp_1p_3)/\text{Area}(\triangle p_1p_2p_3)$ and $\beta_3=\text{Area}(\triangle pp_1p_2)/\text{Area}(\triangle p_1p_2p_3)$. Let $s$ be some scalar quantity sampled on the vertices $p_1, p_2$ and $p_3$ as $s_1, s_2$ and $s_3$ respectively. Then the value of $s$ at any point $p$ inside the triangle is given by $s(p) = \beta_1s_1+\beta_2s_2+\beta_3s_3$.
}
\label{fig:barycenter}
\end{figure}

For a triangle with vertices $p_1=(x_1,y_1)$, $p_2=(x_2,y_2)$ and $p_3=(x_3,y_3)$, the barycentric coordinates $(\beta_1, \beta_2, \beta_3)$ of an arbitrary point $p=(x,y)$ within a triangle are given by 
\begin{align}
    \beta_1 &= \frac{(y_2-y_3)(x-x_3) + (x_3-x_2)(y-y_3)}{(y_2-y_3)(x_1-x_3)+(x_3-x_2)(y_1-y_3)},  \nonumber \\ 
    \beta_2 &= \frac{(y_3-y_1)(x-x_3) + (x_1-x_3)(y-y_3)}{(y_2-y_3)(x_1-x_3)+(x_3-x_2)(y_1-y_3)},  \nonumber \\ 
    \beta_3 &= \frac{(y_1-y_2)(x-x_2) + (x_2-x_1)(y-y_2)}{(y_2-y_3)(x_1-x_3)+(x_3-x_2)(y_1-y_3)}.  \nonumber
\end{align}
We assume that some scalar function $s$ is sampled at the vertices of the triangle with scalar values $s_1$, $s_2$ and $s_3$ at vertices $p_1$, $p_2$ and $p_3$. Then, the scalar value $s$ at an arbitrary point $p=(x,y)$ within the triangle can be determined as
\begin{align}
    s(x,y) &= \beta_1 s_1 + \beta_2 s_2 + \beta_3 s_3.  \nonumber
\end{align}
This function $s(x,y)$ is linear in $x$ and $y$, and can be alternatively written as
\begin{equation}
    s(x,y) = s_x x + s_y y + s_c,  \nonumber
\end{equation}
with 
\begin{align}
    s_x &= \frac{(y_2-y_3)s_1 + (y_3-y_1)s_2 + (y_1-y_2)s_3}{(y_2-y_3)(x_1-x_3)+(x_3-x_2)(y_1-y_3)}  \nonumber \\ 
    s_y &= \frac{(x_3-x_2)s_1 + (x_1-x_3)s_2 + (x_2-x_1)s_3}{(y_2-y_3)(x_1-x_3)+(x_3-x_2)(y_1-y_3)}  \nonumber \\ 
    s_c &= \frac{(x_2 y_3 - x_3 y_2)s_1 + (x_3 y_1 - x_1 y_3)s_2 + (x_1 y_2 - x_2 y_1)s_3}{(y_2-y_3)(x_1-x_3)+(x_3-x_2)(y_1-y_3)}  \nonumber
\end{align}

\subsection{Bivariate quadratic functions and their critical points} \label{section:back_quadratic}
As the anisotropy, as defined in equation~\eqref{eq:anisotropyGeneral}, is a quadratic function we summarize some general facts about quadratic functions and define the notation that we will use later in the paper.

A general quadratic function in two variables can be written as
    $s(x,y) = Ax^2 + Bxy + Cy^2 + Dx + Ey + F$ 
or as quadratic form as 
\begin{equation}
s(x,y) = (x,y)\cdot M \cdot   
          \begin{pmatrix} x \\ y \end{pmatrix}  
         + T \cdot\begin{pmatrix} x \\ y \end{pmatrix} +F 
         \label{eq:generalQuadratic}
\end{equation}
 with 
$M=
  \begin{pmatrix}
    A & B/2 \\
    B/2 & C
  \end{pmatrix}  
$
 and  
$  T=
  \begin{pmatrix}
    D \\
    E
  \end{pmatrix}  
 $.
 The critical point of  the scalar function $s$ is a point $p=(x,y)$ where the gradient $\nabla s(x,y) = 0$ is zero. 
\begin{figure}[ht]
 \centering
 \subcaptionbox[]{\label{fig:ellipticMin}}
    {\includegraphics[width=0.30\linewidth]{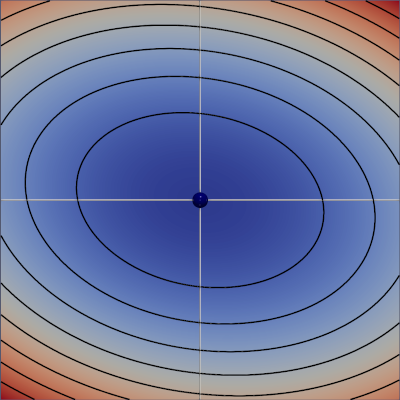}}
 \subcaptionbox[]{\label{fig:ellipticMax}}
    {\includegraphics[width=0.30\linewidth]{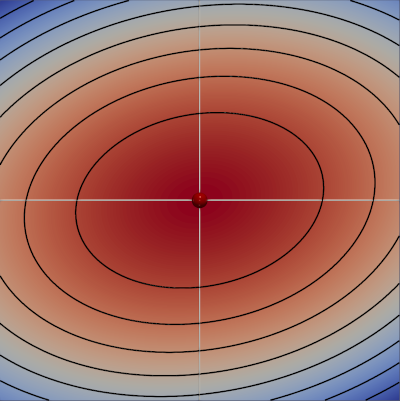}}
 \subcaptionbox[]{\label{fig:hyperbolic}}
    {\includegraphics[width=0.30\linewidth]{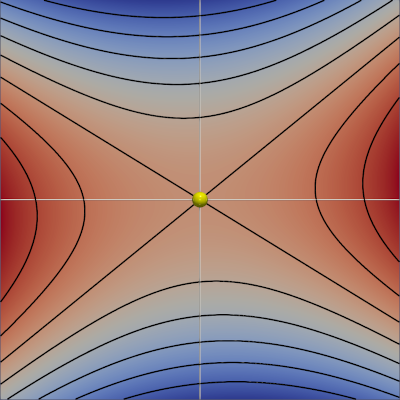}} \\
 \subcaptionbox[]{\label{fig:parallelMin}}
    {\includegraphics[width=0.30\linewidth]{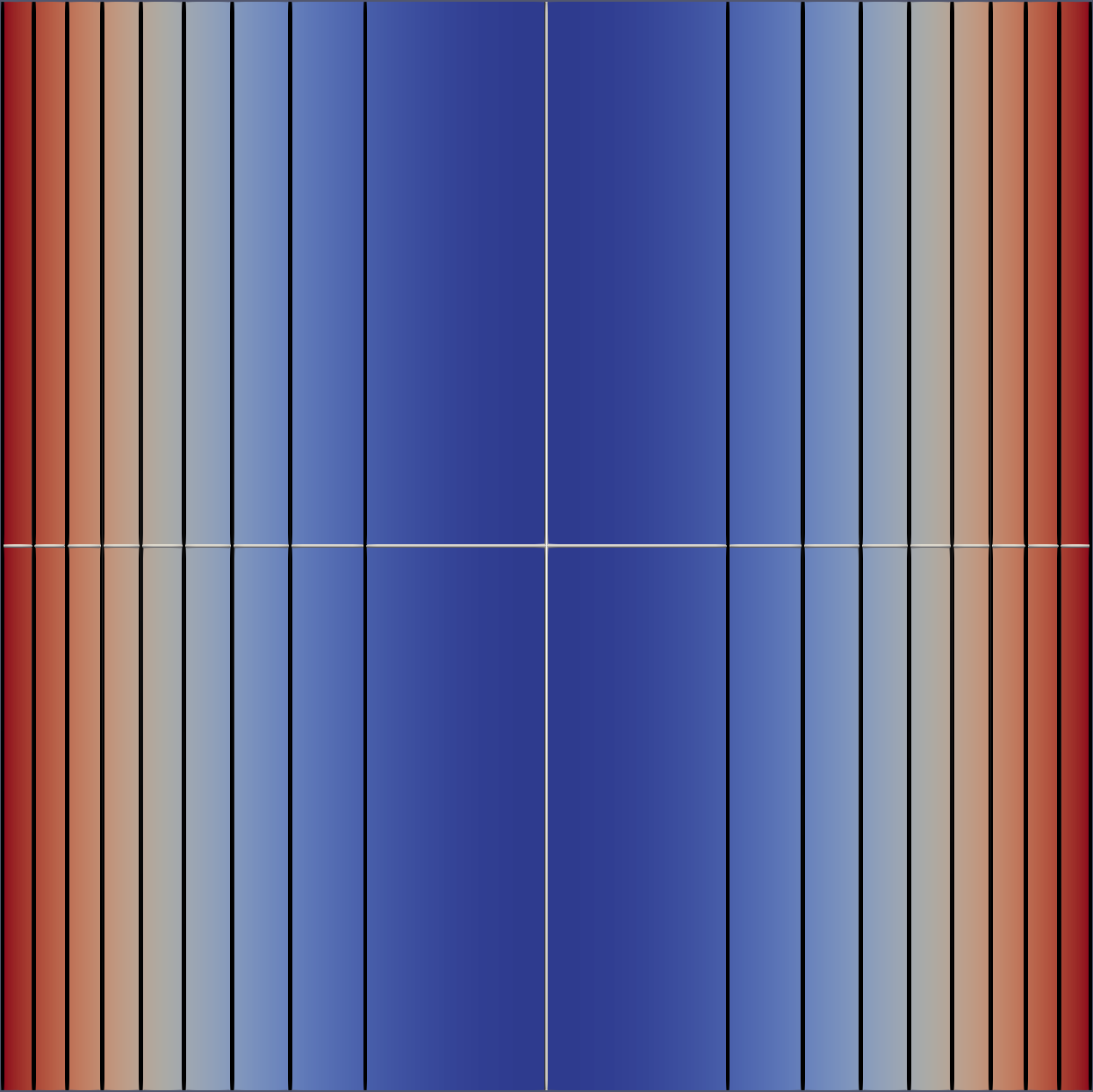}}
 \subcaptionbox[]{\label{fig:parallelMax}}
    {\includegraphics[width=0.30\linewidth]{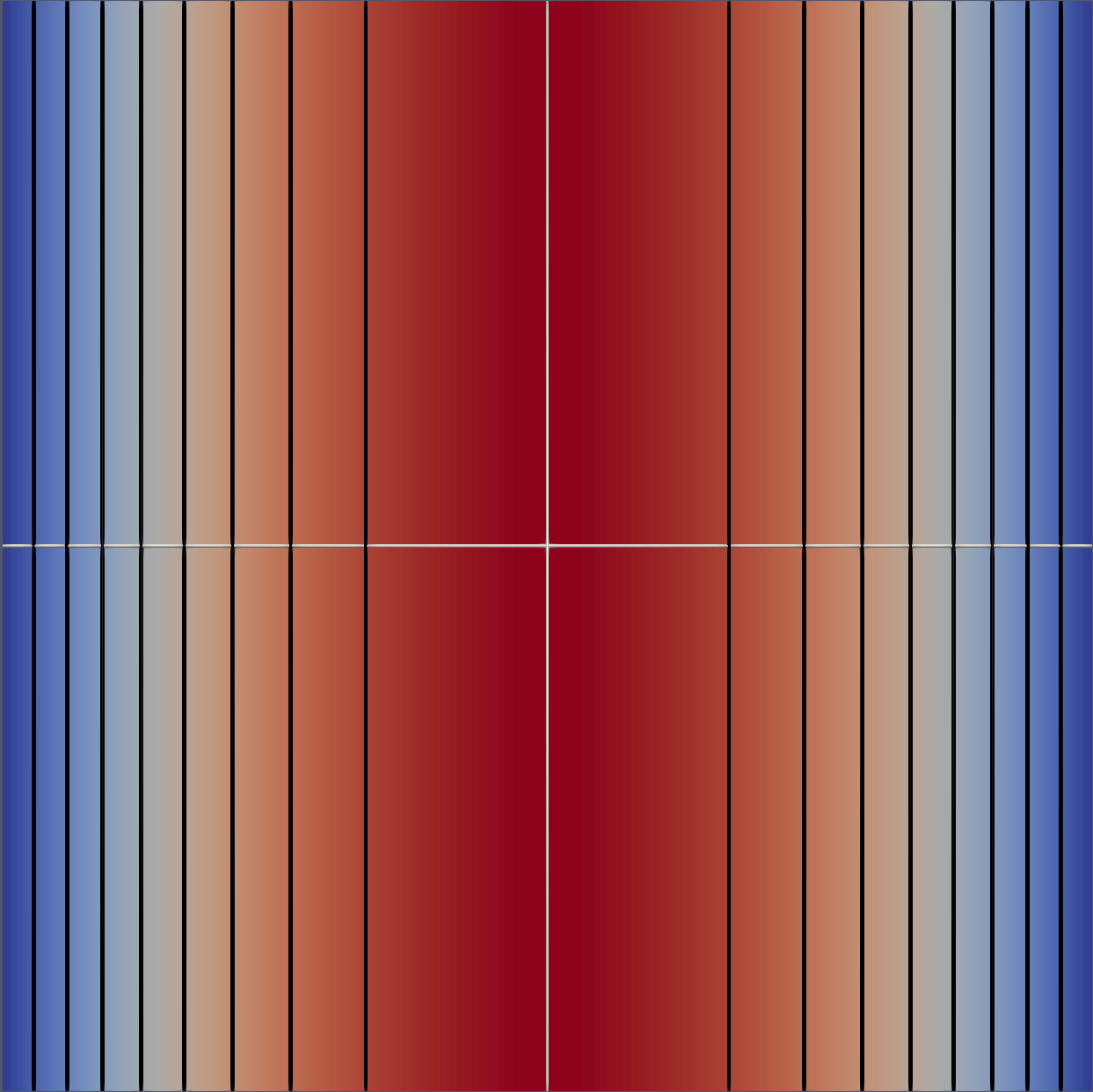}}
 \subcaptionbox[]{\label{fig:parabolic}}
    {\includegraphics[width=0.30\linewidth]{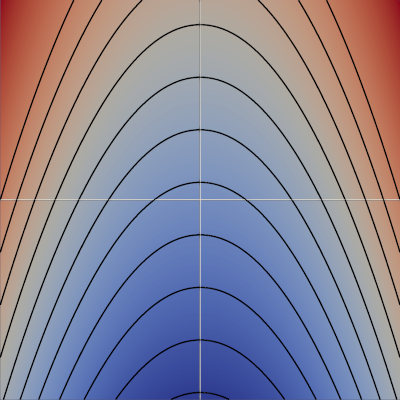}}
  \caption{
  The possible cases for a general quadratic function. 
  \subref{fig:ellipticMin} The critical point is a minimum and the contours are elliptical. 
  \subref{fig:ellipticMax} The critical point is a maximum and the contours are ellipses. 
  \subref{fig:hyperbolic} The critical point is a saddle and the contours are hyperbolas. 
  \subref{fig:parallelMin} The critical point does not exist and the function has a minimum value along a line, while the contours are a pair of parallel lines. 
  \subref{fig:parallelMax} The other case when the contours are a pair of parallel lines, again the critical point does not exist but the function has a maximum value along a line. 
  \subref{fig:parabolic} The case when the critical point does not exist and the contours are parabolic. 
  We show that only the cases \protect\subref{fig:ellipticMin} and \protect\subref{fig:parallelMin} are possible for the quadratic function for tensor anisotropy, the other four cases shown here are not possible.
  }
  \label{fig:quadrics}
\end{figure}
The partial derivatives of $s(x,y)$ with respect to the two variables is:
\begin{align}
    {\partial s}/{\partial x} &= 2Ax + By + D \label{eq:partialDerX} \\ 
    {\partial s}/{\partial y} &= 2Cy + Bx + E \label{eq:partialDerY}
\end{align}

The location of the critical point $p_c = (x_c, y_c)$ of $s$ can be obtained after solving the linear equations $\partial s / \partial x = 0$ and $\partial s / \partial y = 0$ using equations~\eqref{eq:partialDerX} and~\eqref{eq:partialDerY}. The resulting coordinates are
\begin{align}
    x_c = \frac{-2CD + BE}{4AC - B^2}, \qquad 
    y_c = \frac{-2AE + BD}{4AC - B^2}. 
\end{align}

The function $s$ and its critical point can be classified based on the sign of the determinant of the Hessian, $H$ 
\begin{equation}
    H = 4AC - B^2. \label{eq:hessianDet}
\end{equation}

If $H > 0$, then the critical point is either a maximum or minimum and the contours are ellipses, compare Figure~\ref{fig:quadrics}(a,b). The type of the critical point depends on the sign of $A$ and $C$. If $A, C > 0$, then the critical point is a minimum, otherwise it is a maximum. In the other case when $H < 0$, then the critical point is a saddle and the contours are hyperbolas. 

For the case when $H = 0$, there are no critical points and the contours are either parabolas, parallel or coincident lines. The type of the contour when $H = 0$ depends on the invariant $I$ defined as 
\begin{equation}
    I = BDE-AE^2-CD^2. \label{eq:I}
\end{equation}

If $I \neq 0$, the contours are parabolic, compare Figure~\ref{fig:quadrics}(f), otherwise they are pair of parallel lines, compare Figure~\ref{fig:quadrics}(d-c).

%% SECTION 3 Anisotropy
%% ========================================================================= %%
\section{Anisotropy for 2D piece-wise linear tensor fields} \label{section:anisotropy}
In this section, we have a closer look at the anisotropy function for a linear tensor field.
The main observation is in the non-degenerate case, that the anisotropy always has exactly one minimum with a value of zero.
This corresponds to the fact that the tensor field always has exactly one critical point. All contour lines are ellipses. 

Given a triangle with vertices $p_1=(x_1,y_1)$, $p_2=(x_2,y_2)$ and $p_3=(x_3,y_3)$. Let the tensors at these vertices be $T_1$, $T_2$ and $T_3$, respectively. Let the components of the tensors be 
\begin{equation}
T_1=
  \begin{pmatrix}
    e_1 & f_1 \\
    f_1 & g_1 
  \end{pmatrix}
, \quad T_2=
  \begin{pmatrix}
    e_2 & f_2 \\
    f_2 & g_2 
  \end{pmatrix}
, \quad T_3=
  \begin{pmatrix}
    e_3 & f_3 \\
    f_3 & g_3 
  \end{pmatrix}.
\end{equation}

The barycentric coordinates can be used for linear interpolation of the tensor field within the triangle. Tensor $T$ at an arbitrary point $p=(x,y)$ within the triangle can be found as:
\begin{equation}
T(p)= T(x,y) =
  \begin{pmatrix}
    e(x,y) & f(x,y) \\
    f(x,y) & g(x,y) 
  \end{pmatrix}
\end{equation}

As described earlier in Section \ref{section:back_barycenter}, the tensor components $e, f$ and $g$ linearly interpolated within the triangle as can be written as
%After linear interpolation, a linear function is obtained for each tensor component
\begin{align}
    e(x,y) = e_x x + e_y y + e_c, \label{eq:e_linear} \\
    f(x,y) = f_x x + f_y y + f_c, \label{eq:f_linear} \\
    g(x,y) = g_x x + g_y y + g_c. \label{eq:g_linear}
\end{align}
%The complete details of the linear function obtained using barycentric coordinate interpolation for each component is provided in the Appendix~\ref{appendix:barycentric_interpolation}. 

%\subsection{Analysis of anisotropy}
The function, we are interested in is the anisotropy \sqanis, the explicit expression for which can be obtained by substituting the linear expressions for tensor components in equation~\eqref{eq:anisotropyGeneral} yielding
\begin{align}
%    \sqanis(T) = & (e-g)^2+4f^2  \nonumber \\ 
%           = & \big((e_x x + e_y y + e_c) - (g_x x + g_y y + g_c)\big)^2+4(f_x x + f_y y + f_c)^2 \nonumber \\
    \sqanis(T) = & \big((e_x-g_x)^2 + 4f_x^2\big) x^2 + 2\big((e_x-g_x)(e_y-g_y) + 4f_xf_y\big) xy \nonumber \\
             &+ \big((e_y-g_y)^2 + 4f_y^2\big) y^2 + 2\big((e_x-g_x)(e_c-g_c) + 4f_xf_c\big) x\nonumber \\
             &+ 2\big((e_y-g_y)(e_c-g_c) + 4f_yf_c\big) y + \big((e_c-g_c)^2 + 4f_c^2\big). \label{eq:anisQuadratic}
\end{align}
Comparing equation~\eqref{eq:anisQuadratic} with the general quadratic equation~\eqref{eq:generalQuadratic}, we obtain the coefficients of the quadratic function in dependence of the tensor components:
\begin{align}
    A &= \big((e_x-g_x)^2 + 4f_x^2\big) \geq 0 ,  \label{eq:A} \\
    B &= 2\big((e_x-g_x)(e_y-g_y) + 4f_xf_y\big), \label{eq:B} \\
    C &= \big((e_y-g_y)^2 + 4f_y^2\big) \geq 0,   \label{eq:C} \\
    D &= 2\big((e_x-g_x)(e_c-g_c) + 4f_xf_c\big), \label{eq:D} \\
    E &= 2\big((e_y-g_y)(e_c-g_c) + 4f_yf_c\big), \label{eq:E} \\
    F &= \big((e_c-g_c)^2 + 4f_c^2\big).          \label{eq:F}
\end{align}
Substituting the above equations in equation~\eqref{eq:hessianDet}, we derive the determinant of Hessian for the anisotropy to be~(for derivation refer to Appendix~\ref{appendix:anisotropy_analysis}):
\begin{align}
    H  =& 16 \big( f_x(e_y-g_y) - f_y(e_x-g_x) \big)^2 \geq 0. \label{eq:H} 
\end{align}

From equation~\eqref{eq:H}, we can conclude that the contours of anisotropy function are never hyperbolic. In most cases, when $H$ is strictly greater than $0$, the contours are elliptical. Moreover, in that scenario, from equations~\eqref{eq:A} and~\eqref{eq:C}, we can deduce that the type of the critical point is always a minimum. 

Appendix~\ref{appendix:anisotropy_analysis} contains the detailed analysis of \sqanis. We show that the bi-variate quadratic function \sqanis~ is a special function that has elliptical contours or in some cases a pair of parallel lines, ruling out the possibility of hyperbolic or parabolic contours.

%% Sub-SECTION Coordinate Transformation
%% ========================================================================= %%
\subsection{Field normalization using coordinate transformations} \label{sec:transforms}
In the following, we derive a coordinate transformation such that the bivariate quadratic scalar function determined for a triangle has a standard format. This allows for the application of a unified strategy for further analysis.

Equation~\eqref{eq:anisQuadratic} gives the expression of anisotropy in general bivariate quadratic form. We have also established that contours of this function will are elliptical. Therefore we first transform the coordinates such that elliptical contours are centered at the origin and aligned to the axes. This can be achieved by applying a translation and rotation, both rigid-body transformations preserving areas. Then, a transformation is applied such that contours of the bivariate quadratic function become circles rather than ellipses. This can be achieved by applying a non-uniform scaling, a linear transformation which distorts the area by a constant factor given by the determinant of the transformation matrix.

\begin{figure}[ht]
 \centering
 \subcaptionbox[]{\label{fig:origQuadric}}
    {\includegraphics[width=0.24\linewidth]{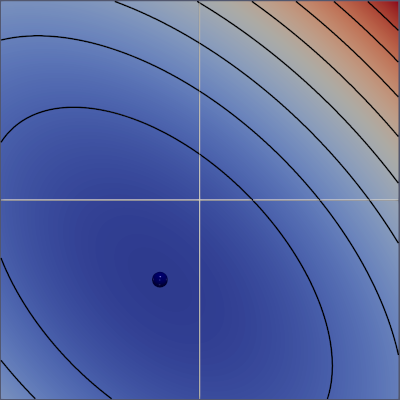}}
 \subcaptionbox[]{\label{fig:afterTrans}}
    {\includegraphics[width=0.24\linewidth]{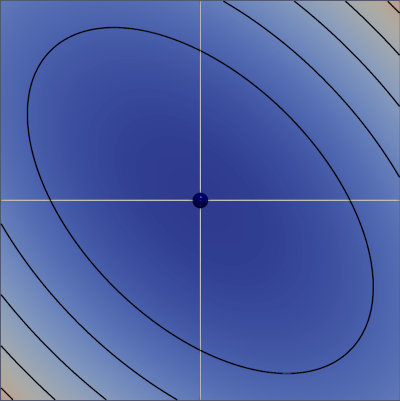}} 
 \subcaptionbox[]{\label{fig:afterRot}}
    {\includegraphics[width=0.24\linewidth]{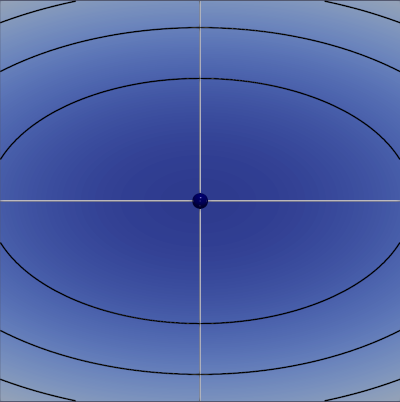}} 
 \subcaptionbox[]{\label{fig:afterScale}}
    {\includegraphics[width=0.24\linewidth]{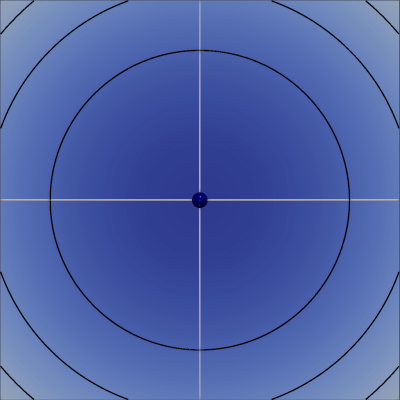}}  
  \caption{
  Field normalization using coordinate transformations. 
  \subref{fig:origQuadric} The original quadratic function, $f(x,y) = x^2+xy+y^2+0.4x+0.5y+0.07$. 
  \subref{fig:afterTrans} The function after applying translation such that the minimum is at the origin, $f(x_t,y_t) = x_t^2+x_ty_t+y_t^2$. 
  \subref{fig:afterRot} After applying rotation such that elliptical contours are axes aligned, $f(x_r,y_r) = 0.5x_r^2+1.5y_r^2$. 
  \subref{fig:afterScale} After scaling such that elliptical contours are transformed into circular contours, $f(x_s,y_s) = x_s^2+y_s^2$.
  }
  \label{fig:coordTransformations}
\end{figure}
%-------derivation in Matrix form -------------
%-------------------------------------------------
After translation such that the minimum $p_c=(x_c, y_c)$ falls into the origin the anisotropy becomes
\begin{equation}
\sqanis(x_t,y_t) = (x_t,y_t)\cdot M \cdot   
          \begin{pmatrix} x_t \\ y_t \end{pmatrix}  
          +F_t 
        \label{eq:translated_with_Ft}
\end{equation}
with the translated coordinates $x_t = x-x_c$, $y_t = y-y_c$ and
     $F_t = F + \frac{(BDE-AE^2-CD^2)}{4AC-B^2}.$
Using equations~\eqref{eq:A}--\eqref{eq:F}, it can be shown that $F_t=0$ for \sqanis. This implies that minimum value of $\sqanis$ is zero at the critical point and this point is a degenerate point in the tensor field.
Equation~\eqref{eq:translated_with_Ft} becomes
\begin{equation}
    \sqanis(x_t,y_t) = (x_t,y_t)\cdot M \cdot   
          \begin{pmatrix} x_t \\ y_t \end{pmatrix}  
    \label{eq:translated} 
\end{equation}

Now, we apply a rotation $O=(EV_1|EV_2)$ to align the elliptic contours with the axis using the eigenvectors $EV_1$ and $EV_2$ of the matrix $M$
\begin{equation}
    \begin{pmatrix} x_r \\ y_r \end{pmatrix}= O\cdot \begin{pmatrix} x_t \\ y_t \end{pmatrix},    
\end{equation}
which results in the diagonal representation of the anisotropy
\begin{equation}
    \sqanis(x_r,y_r) = \lambda_1 x_r^2 + \lambda_2 y_r^2 
\end{equation}
where $\lambda_1$ and $\lambda_2$ are the eigenvalues of $M$.

Lastly, we apply non-uniform scaling to obtain circular contours:
\begin{align}
    \sqanis(x_s,y_s) &= x_s^2 + y_s^2 \label{eq:transformedFn} \quad 
     \text{where}\quad  x_s = \sqrt{\lambda_1} x_r, \quad  y_s = \sqrt{\lambda_2} y_r. 
\end{align}
The area distortion because of this scaling is given by the factor $\sqrt{\lambda_1 \lambda_2}$.

%% Sub-SECTION subdivision
%% ========================================================================= %%
\section{Subdivision in monotonous sub-triangles} \label{section:subdivision}

We consider a triangle with vertices $p_1=(x_1,y_1)$, $p_2=(x_2,y_2)$ and $p_3=(x_3,y_3)$ and the tensors at these vertices are $T_1$, $T_2$ and $T_3$. The tensors are linearly interpolated within the triangle and the anisotropy is a quadratic function. 
For the extraction of contours, the computation of the contour tree as well as for the correct histogram we require piece-wise monotonous behavior inside the triangles, which is given in this chapter. Similar, subdivisions have also been proposed by Dillard{\em~et~al.}~\cite{Dillard2009} and by Nucha{\em~et~al.}~\cite{Nucha2017}.
There are five different cases depending on the location of the global minima and the local minima at the edges. Especially the cases when the minimum $p_c$ lies within the triangle or outside the triangle need a different treatment for the computation of the anisotropy histogram. \\

\begin{figure}[htb]
 \centering
 \subcaptionbox[]{\label{fig:caseFaceCrit}}
    {\includegraphics[width=0.22\linewidth]{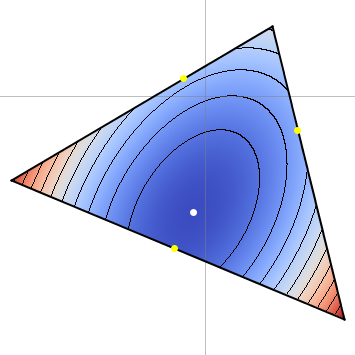}}
    \quad
 \subcaptionbox[]{\label{fig:caseFaceCritDiv}}
    {\includegraphics[width=0.22\linewidth]{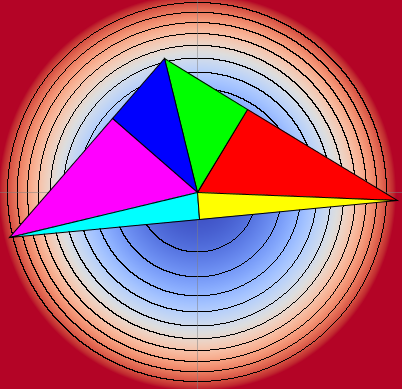}} 
    \quad
 \subcaptionbox[]{\label{fig:caseThreeCrits}}
    {\includegraphics[width=0.22\linewidth]{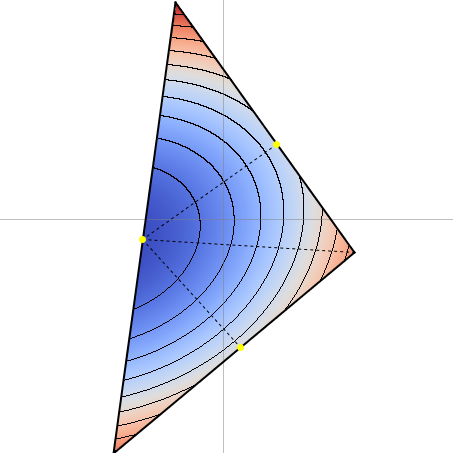}}
    \quad
 \subcaptionbox[]{\label{fig:caseThreeCritsDiv}}
    {\includegraphics[width=0.22\linewidth]{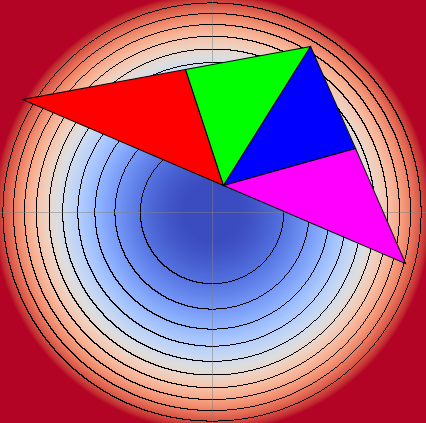}} \\
 \subcaptionbox[]{\label{fig:caseTwoCrits}}
    {\includegraphics[width=0.22\linewidth]{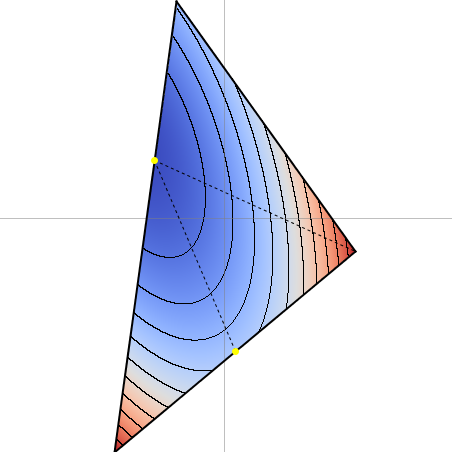}}
    \quad
 \subcaptionbox[]{\label{fig:caseTwoCritsDiv}}
    {\includegraphics[width=0.22\linewidth]{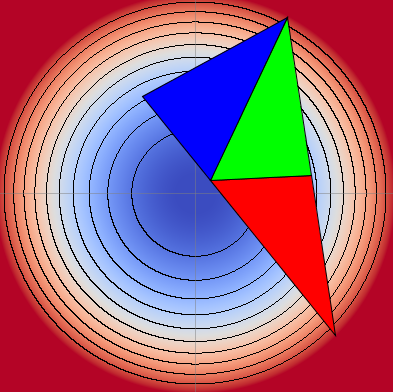}} 
    \quad
 \subcaptionbox[]{\label{fig:caseOneCrit}}
    {\includegraphics[width=0.22\linewidth]{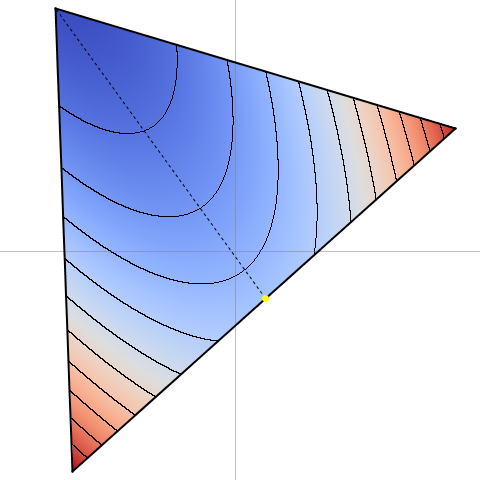}}
    \quad
 \subcaptionbox[]{\label{fig:caseOneCritDiv}}
    {\includegraphics[width=0.22\linewidth]{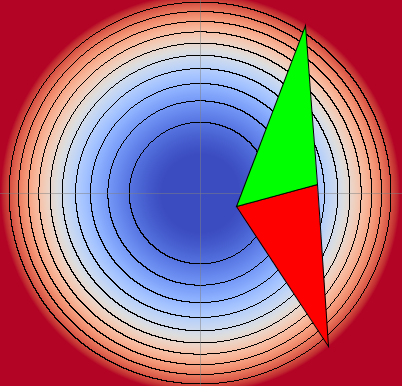}} \\ 
 \subcaptionbox[]{\label{fig:caseNoCrits}}
    {\includegraphics[width=0.22\linewidth]{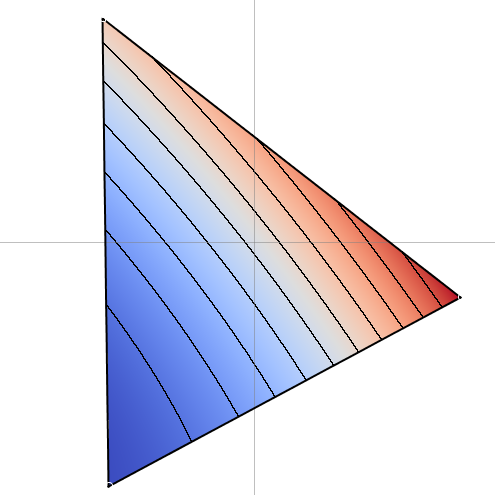}}
    \quad
 \subcaptionbox[]{\label{fig:caseNoCritsDiv}}
    {\includegraphics[width=0.22\linewidth]{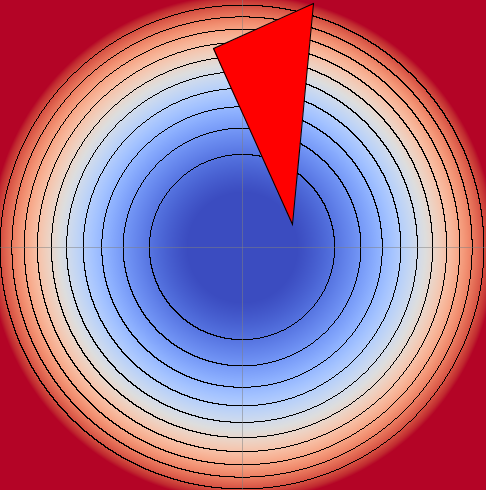}} 
    \quad
 \subcaptionbox[]{\label{fig:caseParallel}}
    {\includegraphics[width=0.22\linewidth]{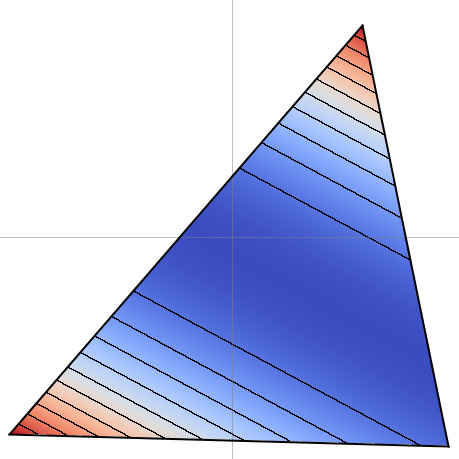}}
    \quad
 \subcaptionbox[]{\label{fig:caseParallelFull}}
    {\includegraphics[width=0.22\linewidth]{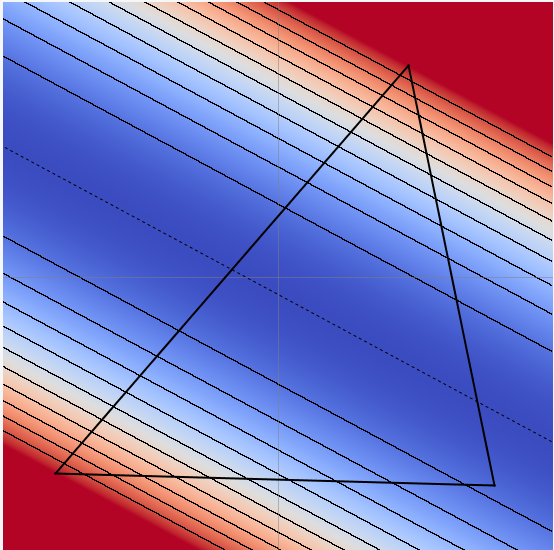}}
  \caption{
  All the possible cases and the corresponding sub-division into monotonous triangles. 
  \subref{fig:caseFaceCrit}, \protect\subref{fig:caseFaceCritDiv} The case when the minimum is inside the triangle. We can subdivide the triangle into six monotonous triangles.
  \subref{fig:caseThreeCrits}, \protect\subref{fig:caseThreeCritsDiv} The case when the minimum is outside the triangle, but the function restricted to the triangle edges has a minimum on all the edges. Four monotonous triangles can be generated.
  \subref{fig:caseTwoCrits}, \protect\subref{fig:caseTwoCritsDiv} The case when two of the edges have minima, three monotonous triangles are generated.
  \subref{fig:caseOneCrit}, \protect\subref{fig:caseOneCritDiv} The case when only one triangle edge has a minimum, two monotonous triangles are created.
  \subref{fig:caseNoCrits}, \protect\subref{fig:caseNoCritsDiv} The case when none of the edges has a minimum, no triangle subdivision is required.
  \subref{fig:caseParallel}, \protect\subref{fig:caseParallelFull} Lastly, the degenerate case when the function does not have a point minimum but has a degenerate minimum along a line. In this case, the contours are a pair of parallel lines. 
  }
  \label{fig:allCases}
\end{figure}

%-------------------------------------%
\begin{description}
\item [\textbf{Case a}] The minimum $p_c$ is within the triangle\\
In this case, the input triangle can be partitioned into six sub-triangles such that \sqanis~ behaves monotonously within the triangle. See Figure~\ref{fig:caseFaceCritDiv}. \\
\item [\textbf{Case b}] The minimum $p_c$ is outside the triangle\\ 
Here, we have four possibilities:
\begin{itemize}
    \item[1.] No triangle edge has an edge minimum. See Figure~\ref{fig:caseNoCrits}.
    \item[2.] One triangle edge has an edge minimum. See Figure~\ref{fig:caseOneCrit}.
    \item[3.] Two triangle edges have edge minima. See Figure~\ref{fig:caseTwoCrits}.
    \item[4.] All three triangle edges have edge minima. See Figure~\ref{fig:caseThreeCrits}.
\end{itemize}
In all these cases, the triangle can be sub-divided into an appropriate number of monotonous triangles. Although these sub-divided triangles are monotonous and look similar to the first case, an important difference is that in general none of the vertices lies at the origin. 
\end{description}

%% Histogram
%% ========================================================================= %%
\section{Computation of the  histogram for \sqanis} \label{section:computation}

In this section, we derive the exact continuous histogram of the anisotropy for linearly interpolated tensor fields.
We use the term histogram here not strictly as it has been originally introduced
but in the sense of the distribution of function values. It can also be considered as the \emph{weighted contour spectrum} using the terminology of Bajaj{\em~et~al.}~\cite{Bajaj1997}
as introduced by Scheidegger{\em~et~al.}~\cite{Scheidegger2008}.
In contrast to previous methods, we approach the problem by first computing the cumulative histogram $CH$ and then derive the histogram from it by  computing its derivative. This makes the exact computation of the histogram much more feasible.
\[CH(\sqanis_0)=Area(\sqanis\le\sqanis_0)  = \sum_{\triangle_i}Area_i(\sqanis\le\sqanis_0)\]
Where $Area_i(\sqanis\le\sqanis_0)$ is the area of the sublevel set in triangle $i$.
Specifically, we consider here the anisotropy with its quadratic behavior, however the method also directly applies for linear fields. 

%--------------------
In our derivation of $Area_i(\sqanis\le\sqanis_0)$ we consider the following setting. Given is a triangle $\triangle ABC$
with vertices $A=p_1=(x_1,y_1)$, $B=p_2=(x_2,y_2)$ and $C=p_3=(x_3,y_3)$ and respective tensors $T_1$, $T_2$ and $T_3$, which are linearly interpolated within the triangle. 
Further we ordered the vertices such that the anisotropy values are $\sqanis_1 < \sqanis_2 < \sqanis_3$. 
To obtain the contribution of the triangle to the cumulative histogram at the value $\sqanis$ we compute the area of the sublevel set at $\sqanis$ within the triangle.
In the following, we assume that anisotropy is monotonous within the triangles resulting from the subdivision introduced in Section~\ref{section:subdivision}.
We further assume that the transformations as described in Section~\ref{sec:transforms} have been applied. This means that the anisotropy has the form as given in equation~\eqref{eq:transformedFn} with a  global minimum $p_c=(0,0)$ at the origin and the level sets that are circles. The area within the transformed triangle is distorted by a constant factor, which can be appropriately multiplied to get the exact areas for the original triangle. We consider two cases depending on whether the vertex $A$ of the triangle vertex lies at the origin or not.

%-------------------------------------%
\subsection*{Case 1: The global minimum $p_c$ lies at one triangle vertex} \label{section:case1}

Let's assume that the global minimum $p_c$ lies in vertex $A=p_1$ of the triangle $ABC$.
Then the shape of the sublevel set for $\sqanis$ can have two different types depending
on whether $\sqanis$ is smaller or larger than $\sqanis_2=r_{AB}^2$. 

\begin{figure}[ht]
 \centering
 \includegraphics[width=0.8\linewidth]{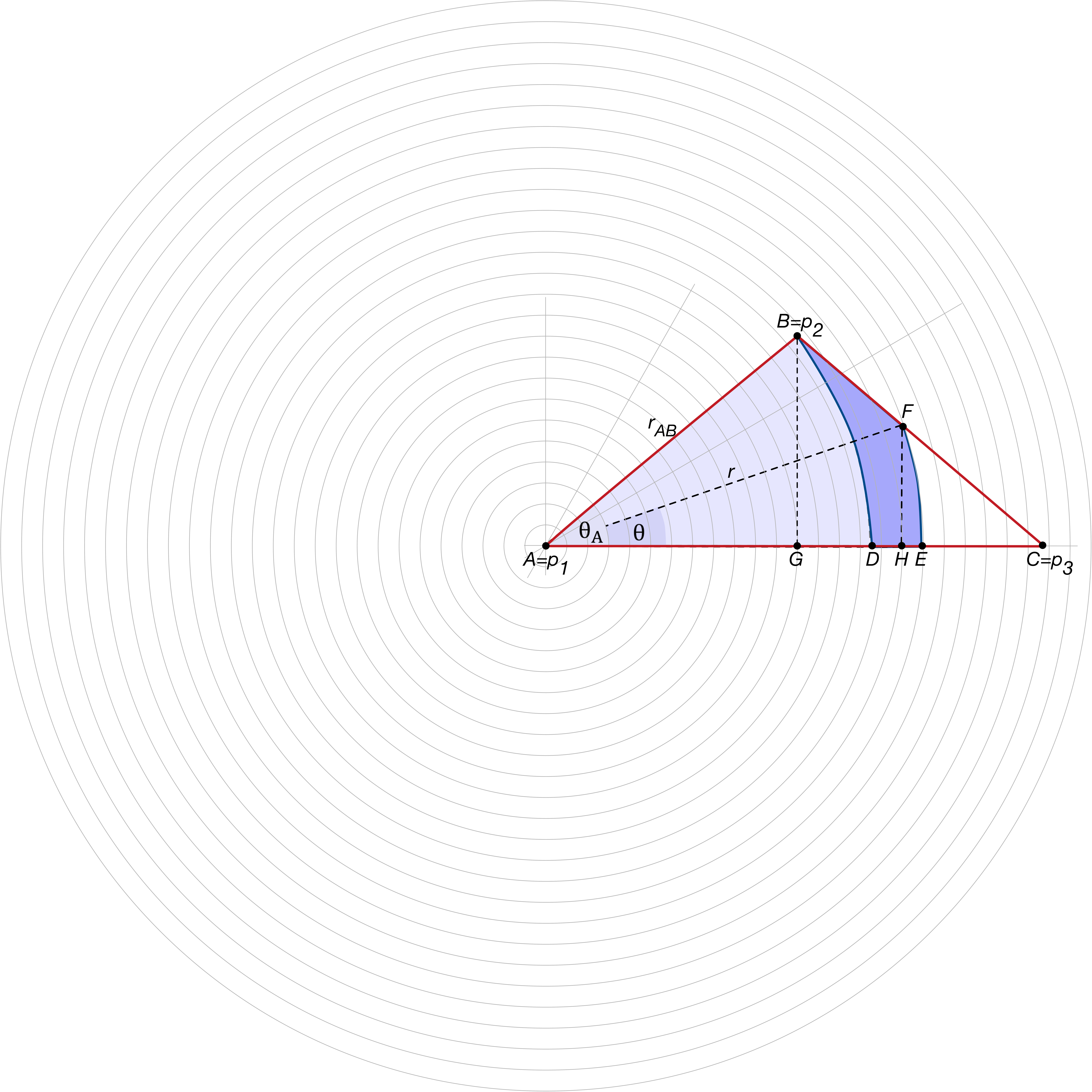}
  \caption{ 
  A triangle with monotonous function behaviour where the global minimum $p_c$ lies on one vertex $A=p_1$. Due to the normalization of the triangle, $p_c$ is at the origin and the contours are circular.
  If $\sqanis\le\sqanis_2$ the sublevel set is a subset of the light blue area. If $\sqanis>\sqanis_2$ the sublevel area is composed of the complete light blue region and the area is highlighted in darker blue.
  }
  \label{fig:redRegion}
\end{figure}

If  $\sqanis\le\sqanis_2$ the sublevel set  is a sector of a circle with radius $r = \sqrt{\sqanis}$ and opening angle $\theta_A$, as shown in Figure~\ref{fig:redRegion}. 
The area of the sublevel set for $\sqanis\in[\sqanis_1 , \sqanis_2]$ is then given by
\begin{align}
    Area(\sqanis) &= \frac{\theta_{A} \sqanis}{2} .
\end{align}
Where $\theta_A$ is angle between the edges $AB$ and $AC$. 

The rate of change of the area is given by
\begin{align}
    \frac{\partial Area(\sqanis)}{\partial \sqanis} &= \frac{\theta_A}{2},
\end{align}
which is a constant not depending on the specific value of $\sqanis$.

If the isovalue \sqanis~is greater than $\sqanis_2$ and less than $\sqanis_3$, the contour is composed of a circular sector and an additional more complex shape $DBFE$ as illustrated in Figure~\ref{fig:redRegion}. 
This region is enclosed by two circular segments $DB$ and $EF$ and two line segments $DE$ and $BF$ and can be computed as
\begin{align}
    Area(BDEF) &= Area(BGHF) + Area(FHE) - Area(BGD) \nonumber \\
      &= \bigg(\frac{(r_{AB} \sin \theta_{A} + r \sin\theta)(r \cos \theta - r_{AB} \cos\theta_{A})}{2}\bigg) \nonumber \\
      &\quad + \bigg(\frac{\theta r^2}{2} - \frac{r^2 \sin\theta \cos\theta}{2}\bigg) \nonumber 
      - \bigg(\frac{\theta_{A} r_{AB}^2}{2} - \frac{r_{AB}^2 \sin\theta_{A} \cos\theta_{A}}{2} \bigg) \nonumber \\
      &= \frac{\theta r^2 - \theta_{A} r_{AB}^2}{2} + r r_{AB} \bigg(\frac{\sin\theta_{A} \cos\theta - \cos\theta_{A} \sin\theta}{2}\bigg). \nonumber 
\end{align}
With $r^2 = \sqanis$ and $r_{AB}^2 = \sqanis_2$ this results in 
\begin{align}
    Area(BDEF) &= \frac{\theta \sqanis - \theta_{A} \sqanis_2}{2} + \sqrt{\sqanis \sqanis_2} \frac{\sin(\theta_{A} - \theta)}{2} .     \label{eq:area}
\end{align}
In the expression above, the given value of \sqanis~determines the values of $\theta$.

\subsection*{Case~2: The global minimum $p_c$ does not lie at any triangle vertex} \label{section:case2}
\begin{figure}[ht]
 \centering
 \includegraphics[width=0.8\linewidth]{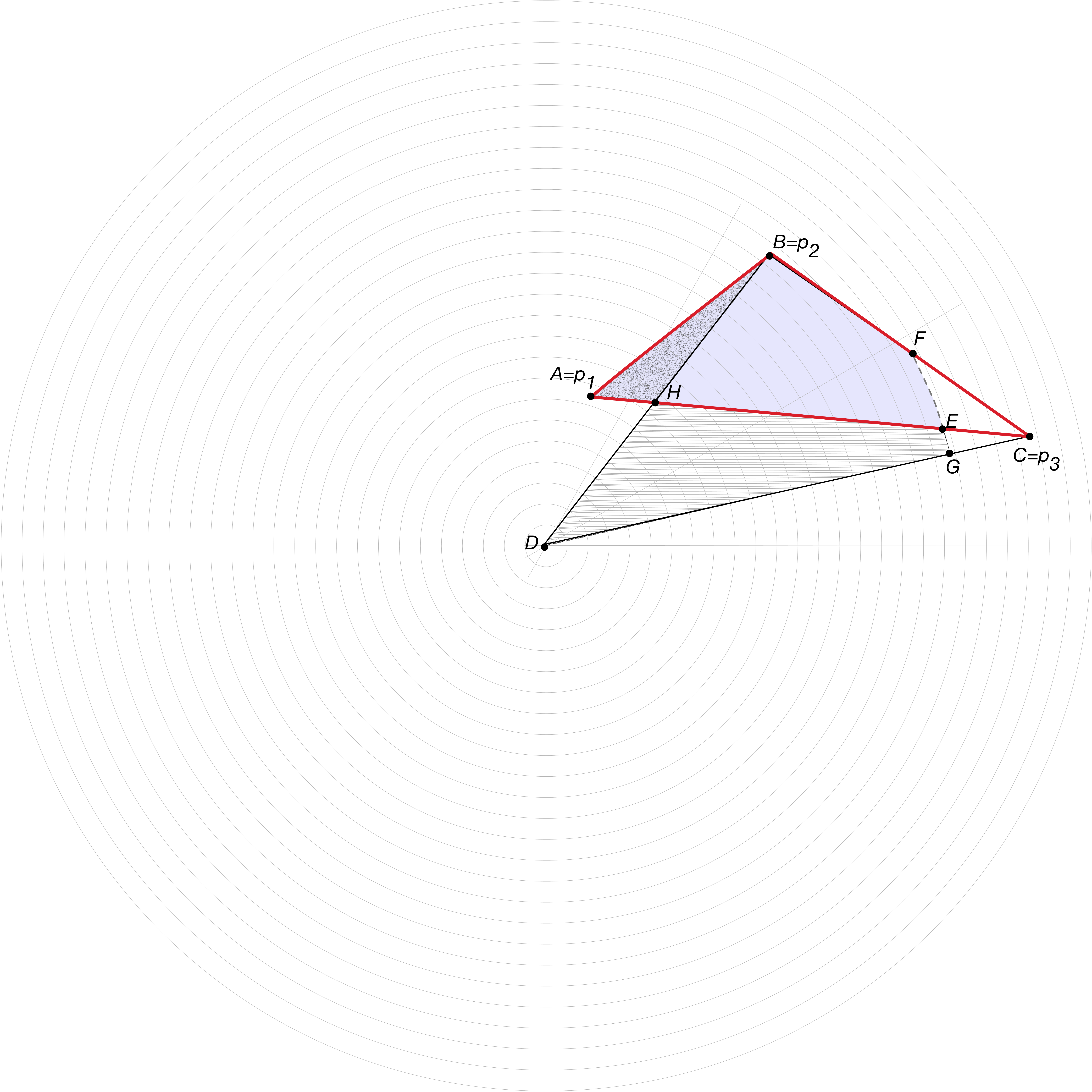}
  \caption{
  A triangle $\triangle ABC$ with monotonous function behaviour where the global minimum $p_c$ does not lie on any vertex.
  }
  \label{fig:area_case2}
\end{figure}
These triangles are the more generic case, we still assume that the anisotropy is monotonous inside the triangle and the vertices such that the anisotropy values are $\sqanis_1 < \sqanis_2 < \sqanis_3$.
The triangles look similar to Case~1, but none of the vertices lie at the origin, compare Figure~\ref{fig:area_case2}. 
To compute the area of a sublevel set $ABFE$ within these triangles, we considering the sublevel sets in two triangles where we can use the algorithm of Case~1. In Figure~\ref{fig:area_case2} they are the triangles $\triangle DBC$ and $\triangle DHC$.  
For the final area, we have in addition to consider the area of triangle $\triangle AHB$
which has to be added or subtracted depending on the exact position of $A$.
\begin{align}
    Area(ABFE) &= Area(DBFG)-Area(DHEG)\pm Area(\triangle AHB)  \label{eq:area-case2}
\end{align}

%%Implementation
%% ========================================================================= %%
%\begin{figure}[htb]
%\begin{minipage}{.85\linewidth}
\begin{algorithm}
\SetAlgoLined
\SetKwFunction{GetSubLevelSetArea}{GetSubLevelSetArea}\SetKwFunction{Range}{Range}
\SetKwData{MT}{MT}\SetKwData{Mesh}{M}\SetKwData{B}{B}\SetKwData{DT}{DT}\SetKwData{CH}{CH}\SetKwData{Tri}{t}
\KwData{A tensor mesh \Mesh, and histogram resolution \B}
\KwResult{Cumulative histogram \CH}
\BlankLine
Initialize histogram \CH to zeros\;
Initialize monotonous triangle list \MT to be empty\;
Initialize degenerate triangle list \DT to be empty\;
\ForEach(in parallel){triangle \Tri in \Mesh}{
    Compute the quadratic function coefficients $A,B,C,D,E,F$\;
    $H = 4AC - B^2$\;
    \eIf{$H>0$}{
        Compute the location of minimum $p_c=(x_c,y_c)$\;
        Compute the number and locations of edge minima\;
        \eIf{$p_c$ is inside \Tri}{
            Generate six monotonous triangles from \Tri and add to \MT\;
        }{
            \Switch{Number of edge minima}{
                \uCase{0}{
                    Add \Tri to \MT\;
                }
                \uCase{1}{
                    Add an edge from edge minimum to opposite triangle vertex\;
                    This divides \Tri in two monotonous triangles\;
                    Add the two monotonous triangles to \MT\;
                }
                \uCase{2}{
                    Add an edge between the two edge minima\;
                    Add an edge from lowest valued edge minimum to opposite vertex\;
                    These two edges divide \Tri into three monotonous triangles\;
                    Add the monotonous triangles to \MT\;
                }
                \Case{3}{
                    Find the lowest valued edge minimum $p_m$\;
                    Add edges between $p_m$ and the two other edge minima\;
                    Add an edge from lowest valued edge minimum to opposite vertex\;
                    These three edges divides \Tri in four monotonous triangles\;
                    Add the monotonous triangles to \MT\;
                }
            }
        }
    }{
        Add \Tri to \DT\;
    }
}
\For(in parallel){$i \leftarrow 1$ to \B}{
    $\sqanis \leftarrow \Range(\MT)/i$ \;
    \CH$[i] \leftarrow 0$\;
    \ForEach(in parallel){monotonous triangle \Tri in \MT}{
        \CH$[i] \leftarrow \CH[i] +$ \GetSubLevelSetArea(\Tri, \sqanis)\;
    }
    \ForEach(in parallel){degenerate triangle $t$ in \DT}{
        \CH$[i] \leftarrow \CH[i] +$ \GetSubLevelSetArea(\Tri, \sqanis)\;
    }
}
\KwRet \CH\;
\caption{Compute cumulative histogram}
\label{algo}
\end{algorithm}
%\end{minipage}
%\end{figure}

\subsection{A brief note on implementation} \label{section:implementation}
We compute the cumulative histogram of anisotropy by computing the sublevel set area. The detailed algorithm is provided in Algorithm~\ref{algo}. For a given tensor mesh, we pre-process all the triangles to determine the coefficients of the quadratic function for anisotropy within the triangle, compare Section~\ref{section:anisotropy}. Each triangle is appropriately subdivided into monotonous triangles, compare Section~\ref{section:subdivision}. Then for each anisotropy value in the histogram bin, we compute the sublevel set area by adding up the sublevel set area in all monotonous triangles. We also handle triangles with degenerate minimum appropriately. 

\subsubsection*{Parallelization} \label{section:parallelization} The algorithm described in Algorithm~\ref{algo} has lots of options for parallelization. The loops at Lines 4 and 38 are embarrassingly parallel because each triangle in the mesh can be pre-processed in parallel. Similarly, each bin in the histogram can be computed in parallel. Further, with atomic add operations, the computation of the sublevel set area for anisotropy value \sqanis~ corresponding to a particular histogram bin can be parallelized over the set of monotonous triangles. These are the loops at Lines 41 and 44 in Algorithm~\ref{algo}.

\subsubsection*{Numerical issues} \label{section:numerical_issues}
Since there are a few floating-point operations and checks involved in the Algorithm~\ref{algo}, floating-point errors can be introduced which may propagate to the output resulting in a wrong histogram. We handle these errors by introducing reasonable checks, for example making sure that the cumulative histogram thus obtained is always a monotonically increasing function and sums up to the total area of the domain for the largest value of anisotropy. However, a deeper study of the errors and more robust computation is left for future work. We are confident that using a multi-precision library for computation will remove the floating-point errors.

%%Results
%% ========================================================================= %%
\section{Results} \label{section:results}
We apply the proposed approach of computing anisotropy histograms to three different case studies. We show results for synthetic, simulation and experimental data.
For all the case studies we compute the continuous histograms for three different interpolations of the anisotropy.
\begin{itemize}
\item \textbf{Interpolation [a]}: Linear interpolation of the anisotropy within the original triangulation.
\item \textbf{Interpolation [b]}: Linear interpolation of the anisotropy within the sub-divided monotonous  triangles, compare Section~\ref{section:subdivision}.
\item \textbf{Interpolation [c]}: Anisotropy based on the linear interpolation of the tensor components, resulting in a quadratic behavior or the anisotropy, compare Section~\ref{section:anisotropy}.
\end{itemize}
The resulting anisotropy fields are directly shown using a color map where dark blue refers to an anisotropy value of zero and red assigned to the maximum value. We also show a set of contours in these images as white lines. In addition, we computed the join tree for all cases, which are always the same for interpolation [b] and [c]. Finally, we computed the continuous histogram where the results for one data set are plotted in one image, interpolation [a] displayed as a red curve, interpolation [b] as a green curve and interpolation [c] as a black curve.

%--------------------------------------------------------
\subsection{Synthetic data} \label{section:synthetic_data}
\begin{figure}[ht]
 \centering
 \begin{tabular}{ccc}
    \includegraphics[height=3.4cm]{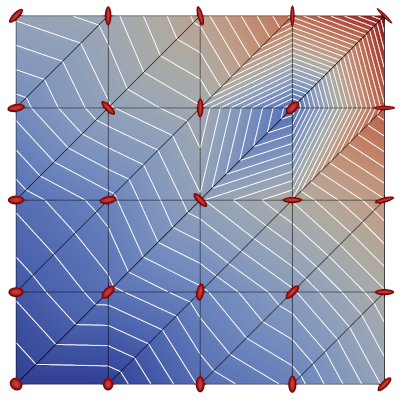} &%\quad%
    \includegraphics[height=3.4cm]{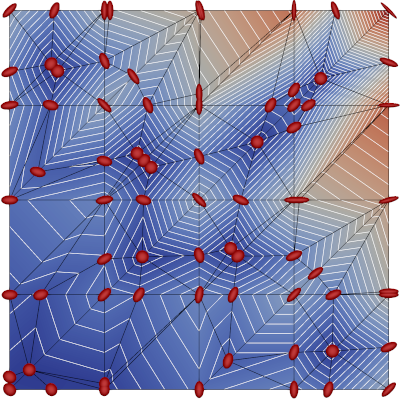} &%\quad% 
    \includegraphics[height=3.4cm]{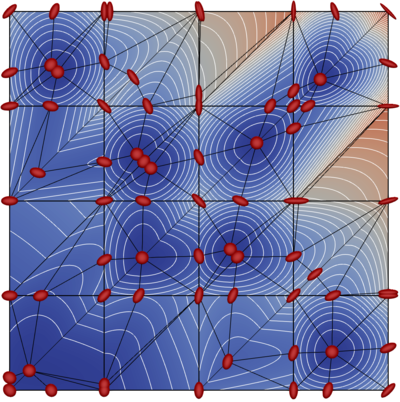}  
    \includegraphics[height=3.0cm]{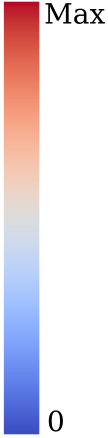}\\%
    \includegraphics[height=3.4cm]{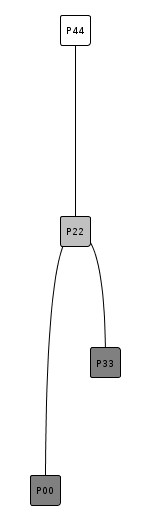}& % \quad%
    \includegraphics[height=3.4cm]{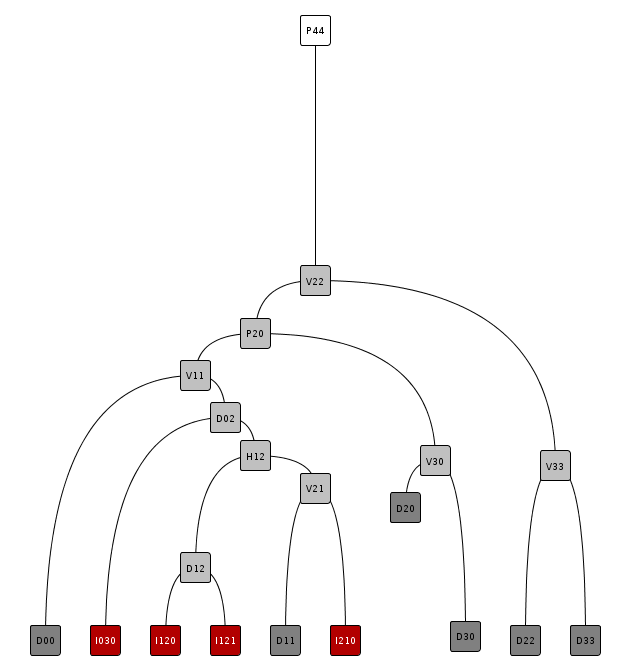} &   %\quad% 
    \includegraphics[height=3.4cm]{mergeTree0.png} \\%
    ~[a] & [b] & [c]
 \end{tabular}
 \caption{
    Comparison of the behavior of the anisotropy field in a synthetic dataset. The anisotropy is linearly interpolated within the original mesh [a], the subdivided monotonous triangles [b]. The correct interpolation of anisotropy assuming linear interpolation of tensor components [c]. The upper row shows the resulting field as a color-map superimposed with contour lines. The ellipses in the vertices represent the tensors defining the field. The second row shows the respective join trees for the three anisotropy fields. The red square corresponds to the zeros inside of the original triangles. All zero-leaves in the tree represent degenerate points of the tensor field topology.
  }
  \label{fig:syn_studyA}
\end{figure}

The first example is a synthetic data set where tensors with user specified tensor components are placed at grid locations of a 5x5 grid. 
This grid is triangulated to provide a mesh with 32 triangles.
This simple example is well suited to demonstrate the differences between the accurate histogram [c], and the methods utilizing a linear approximation of the anisotropy on the original mesh [a], and on the subdivided mesh with monotonous triangles [b], as shown in Figure~\ref{fig:syn_studyA}. 
It can immediately be seen that for interpolation [a] the topology of the contours is not correct. While the exact contours differ between interpolation [b] and interpolation [c], they have the same topology.
This is also reflected in the corresponding join trees, which are the same for both fields Figure~\ref{fig:syn_studyA}. 

\begin{figure}[ht]
 \centering
    \includegraphics[height=2.9cm]{step0_orig.png}%} %
    \includegraphics[height=2.9cm]{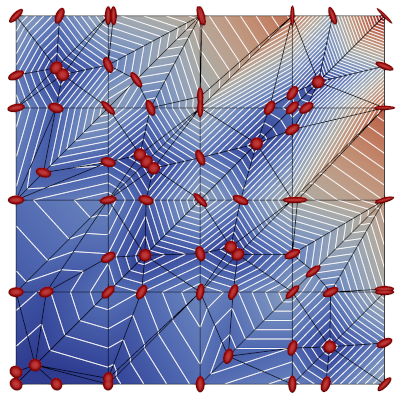}%} % 
    \includegraphics[height=2.9cm]{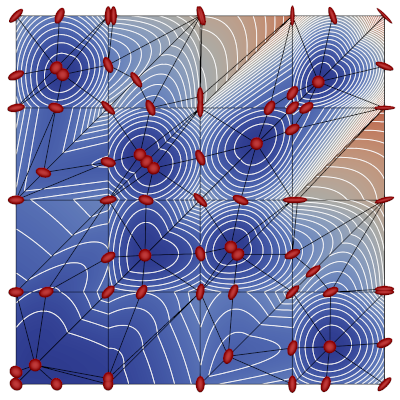}%} %
    \includegraphics[height=2.9cm]{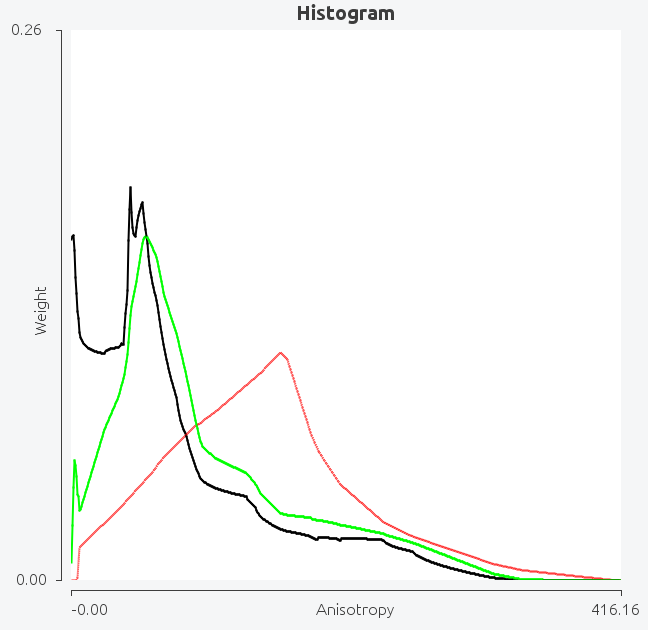} \\%
    \includegraphics[height=2.9cm]{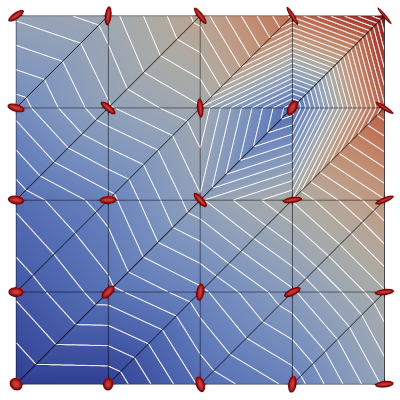}%} %
    \includegraphics[height=2.9cm]{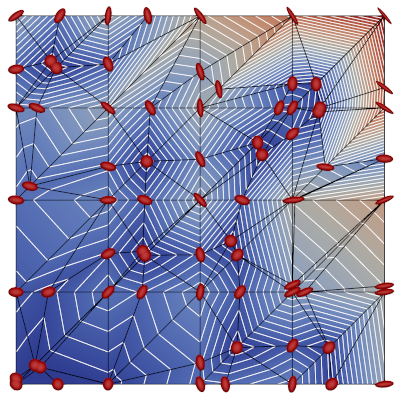}%} % 
    \includegraphics[height=2.9cm]{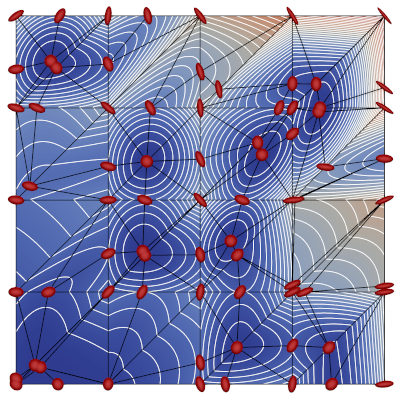}%} %
    \includegraphics[height=2.9cm]{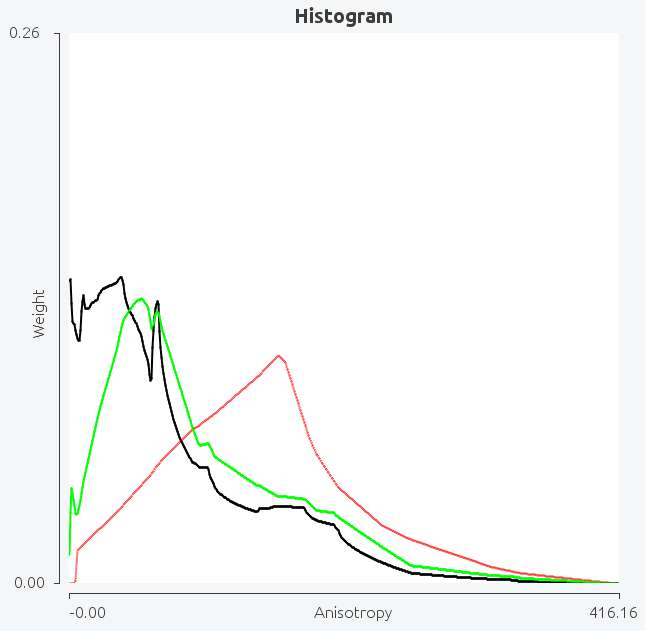} \\%
    \includegraphics[height=2.9cm]{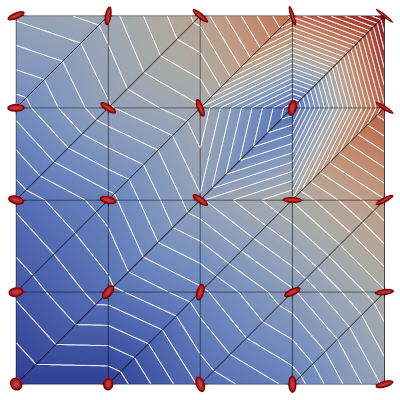}%} %
    \includegraphics[height=2.9cm]{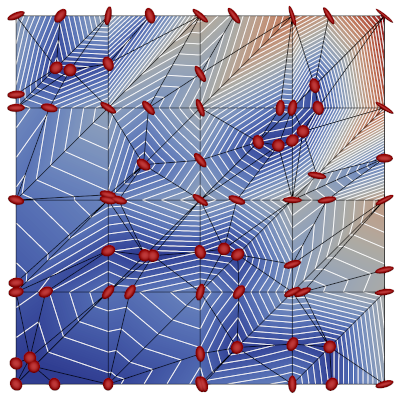}%} % 
    \includegraphics[height=2.9cm]{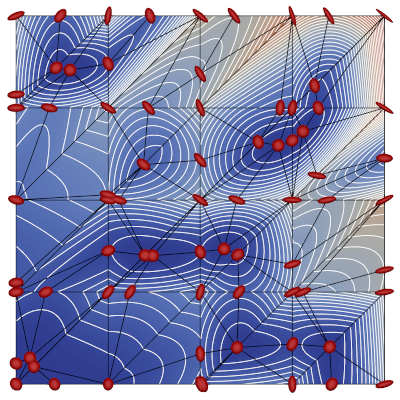}%} %
    \includegraphics[height=2.9cm]{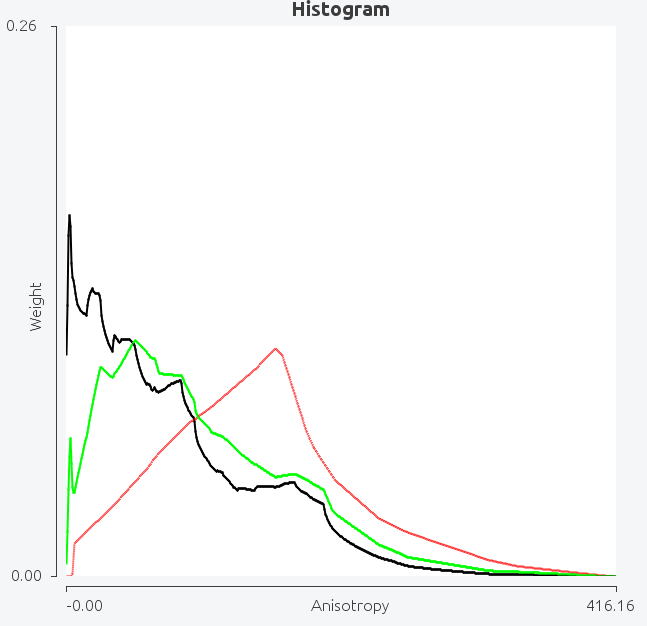} \\%
 \subcaptionbox[]{\label{fig:step9_orig}}%
    {\includegraphics[height=2.9cm]{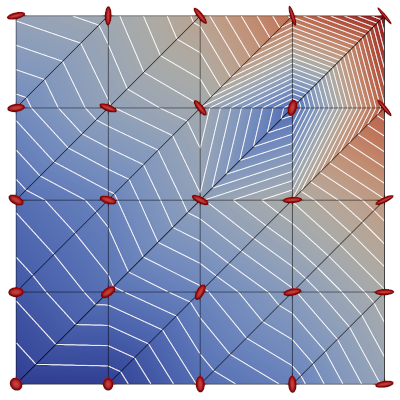}} %
 \subcaptionbox[]{\label{fig:step9_mono}}%
    {\includegraphics[height=2.9cm]{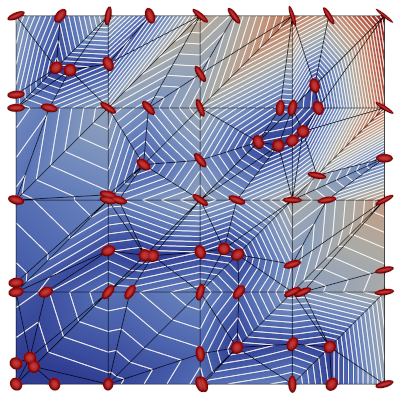}} % 
 \subcaptionbox[]{\label{fig:step9_correct}}%
    {\includegraphics[height=2.9cm]{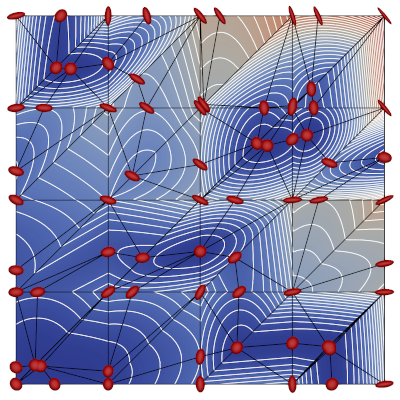}} %
 \subcaptionbox[]{\label{fig:hist9}}%
    {\includegraphics[height=2.9cm]{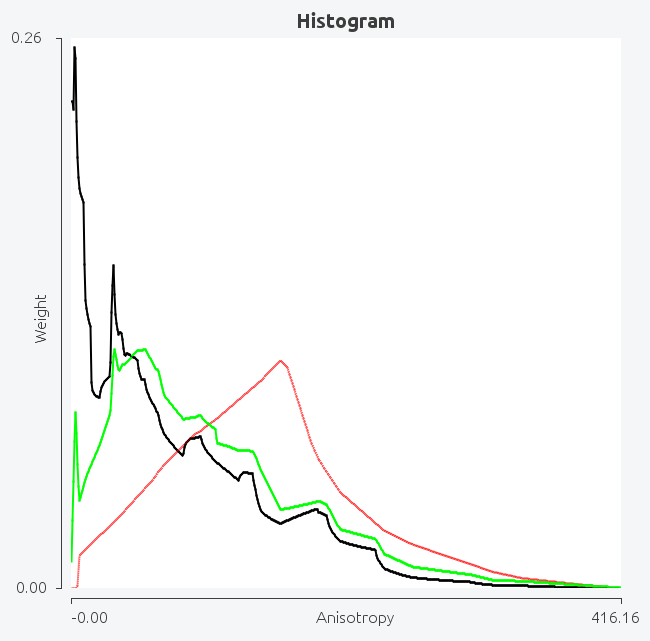}} %
 \caption{Random directions tensor dataset by constant eigenvalues. 
    \subref{fig:step9_orig} The original mesh with anisotropy computed at the mesh vertices and linearly interpolated within the triangles.
    \subref{fig:step9_mono} The subdivided mesh with monotonous triangles. Anisotropy is linearly interpolated within the monotonous triangles.
    \subref{fig:step9_correct} The correct interpolation of anisotropy under linear interpolation of tensor components. 
    \subref{fig:hist9} summarizes all histograms with red curves for interpolation [a], green curves for interpolation [b] and black curves for interpolation [c].
  }
  \label{fig:ensemble_study}
\end{figure}

\begin{figure}[ht]
 \centering
 \begin{tabular}{cccc}
    \includegraphics[height=2.7cm]{mergeTree0.png} & %
    \includegraphics[height=2.7cm]{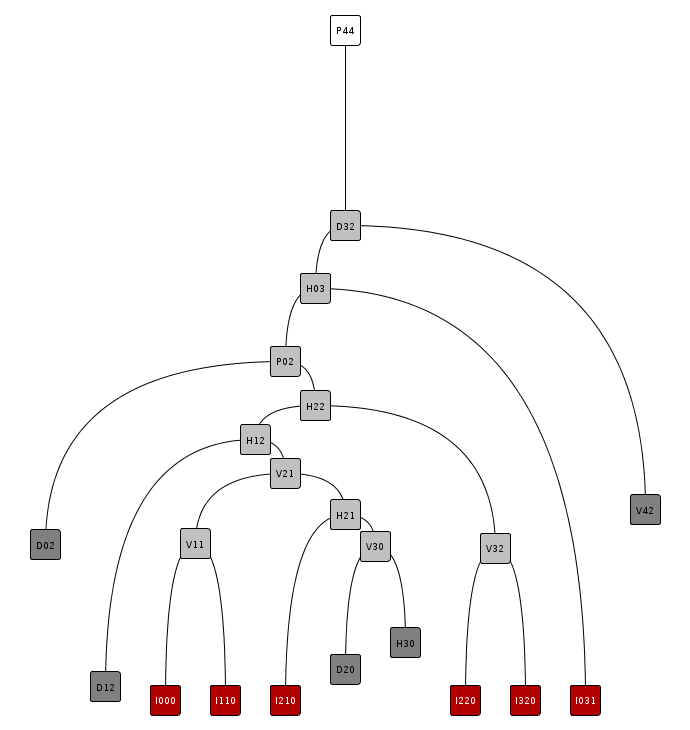} & %
    \includegraphics[height=2.7cm]{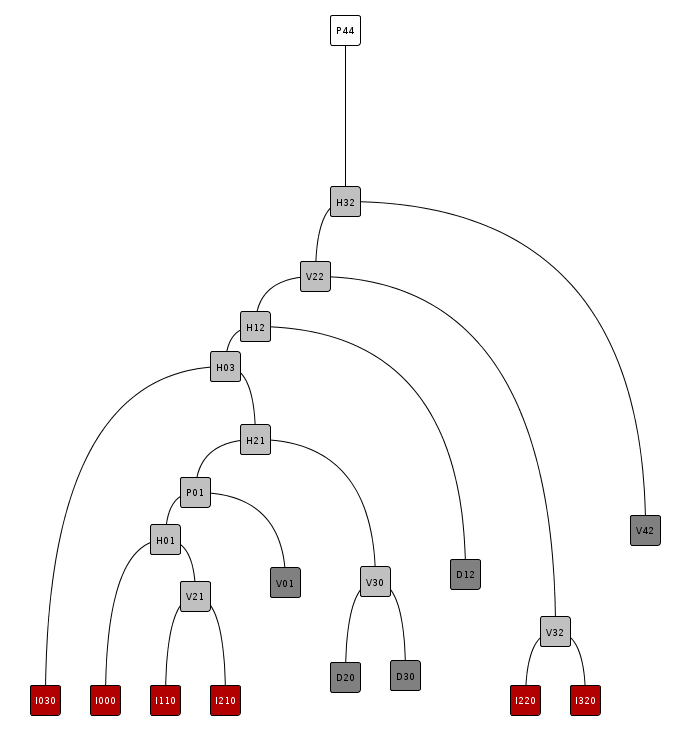} &%
    \includegraphics[height=2.7cm]{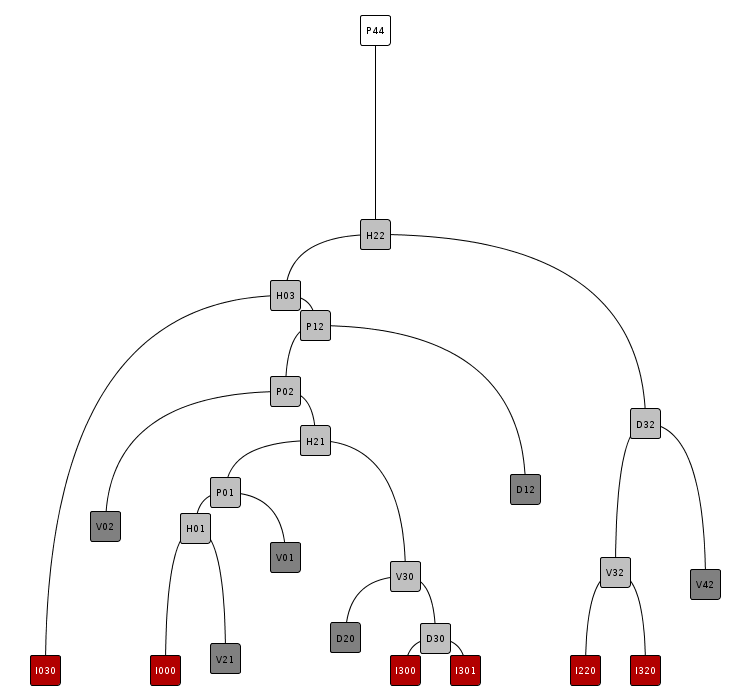} \\%
(1) & (2) &(3) & (4)
\end{tabular}
 \caption{Ensemble of four tensor datasets (1)-(4) generated by random variation of the eigenvector directions at the vertices. 
 The join trees summarize the hierarchical structure of the minima in the anisotropy field
 and the degenerate points in the tensor filed topology. Degenerate points inside the triangles are marked as red nodes in the tree.  
  }
  \label{fig:ensemble_study_trees}
\end{figure}

In the next step, we randomly perturb the directions of the tensors at the vertices without changing their eigenvalues to generate an ensemble of four tensor fields. Note that since we do not change the eigenvalues, the anisotropy at the vertices of the mesh remains unchanged after perturbation. 
Hence, for interpolation [a] 
all the fields, the contours, the join trees, and their histograms  will be the same as evident from Figures~\ref{fig:step9_orig}.
However, if we use correct quadratic function for interpolation [c] of anisotropy, we clearly observe the differences within the ensemble members as shown in Figures~\ref{fig:step9_correct}.
Similarly, the histograms for the ensemble members are different as shown by black curves in the plots in Figures~\ref{fig:hist9}. 
While for interpolation [b], the subdivision into monotonous triangles helps in identifying the differences, Figure~\ref{fig:step9_mono}, the histograms are still not accurate and have a bias toward larger anisotropy values, Figures~\ref{fig:hist9}. 
The respective contour trees are shown in Figure~\ref{fig:ensemble_study_trees}. The contour tree for the interpolation [a] is the same for all 4 fields and already given in  Figure~\ref{fig:step9_orig}, so we only show the trees for the sub-division in monotonous triangles. 
The degenerate points of the tensor field where the anisotropy is zero appear also as minima in the join tree and are highlighted in red. The join trees provide an overview of the possible cancellations of degenerate points and thus their stability~\cite{Wang2017}. 
Comparing the trees it can be seen that the four data sets vary significantly for their topological structure and stability of its degenerate points. The join tree for the linear anisotropy, Figure~\ref{fig:step9_orig}, has no zero and is thus not consistent with the direction fields given by the tensors.

\subsection{Simulation data} 
\begin{figure}
    \centering
    \subcaptionbox[]{\label{fig:two_point_3d}}
    {\includegraphics[height=3.2cm]{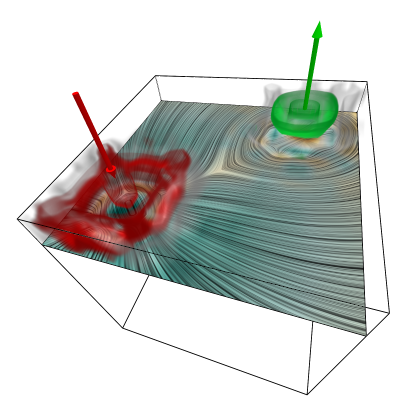}}\quad
    \subcaptionbox[]{\label{fig:two_point_slice}}
    {\includegraphics[height=3.0cm]{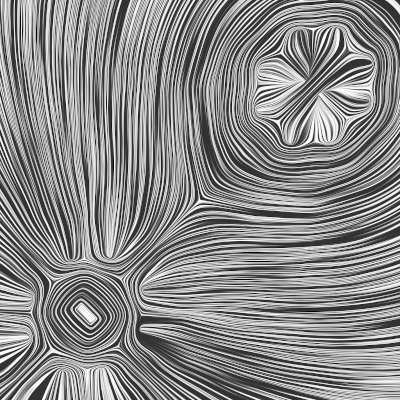}}   \quad \subcaptionbox[]{\label{fig:two_point_slice_select}}
    {\includegraphics[height=3.0cm]{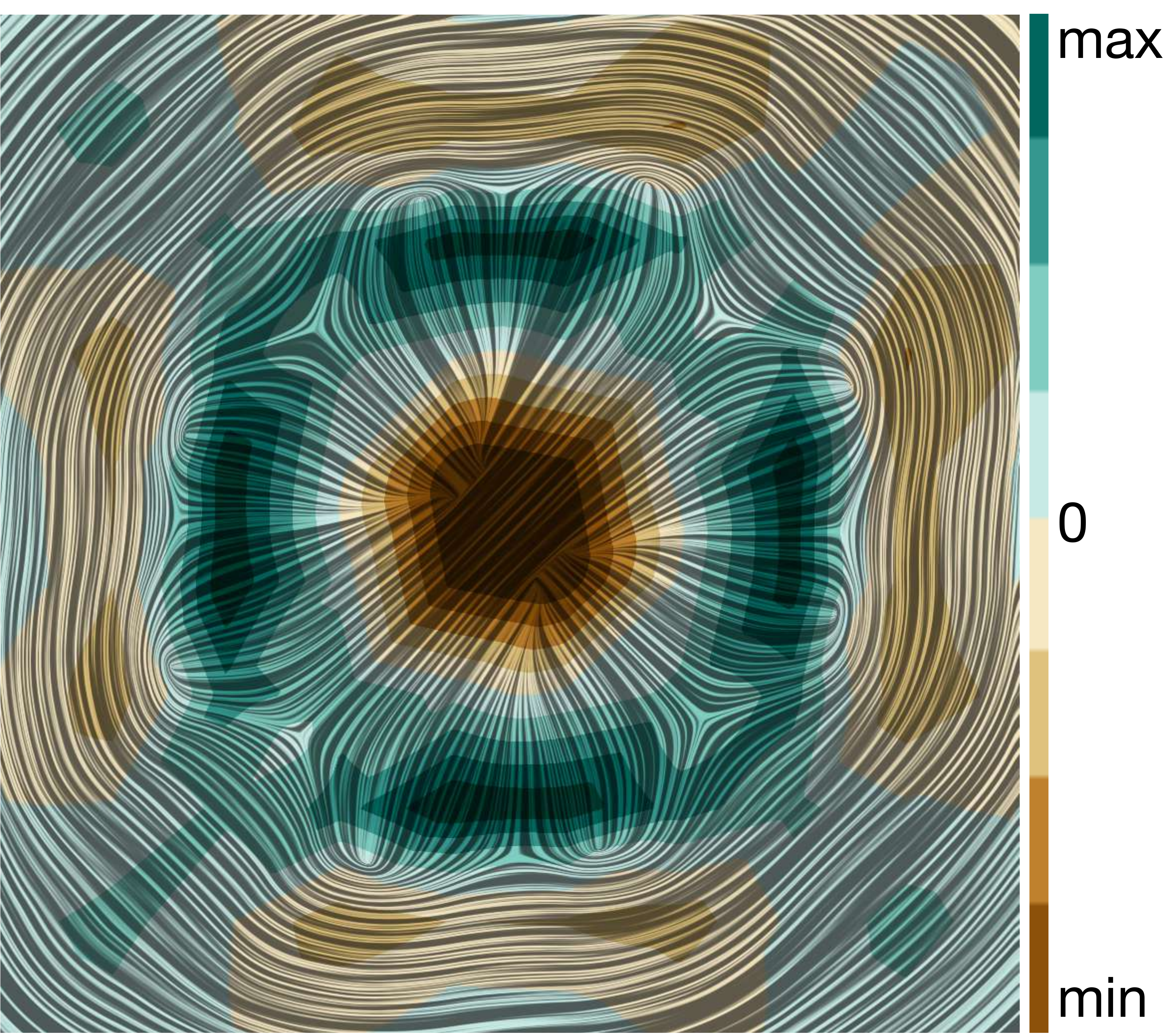}}
    \caption{
    Two-point load data: simulation of the stresses inside a solid block applying two forces.
    \subref{fig:two_point_3d} shows a schematic of the set-up of the simulation, 
    \subref{fig:two_point_slice} shows the eigenvector directions in one slice,
    \subref{fig:two_point_slice_select} is a close up with color showing the signed anisotropy using bi-linear interpolation on the original mesh.
    }
    \label{fig:twoPoint_data}
\end{figure}

\begin{figure}[hb]
 \centering
 \subcaptionbox[]{\label{fig:twoPoint_orig}}%
    {\includegraphics[height=3.0cm]{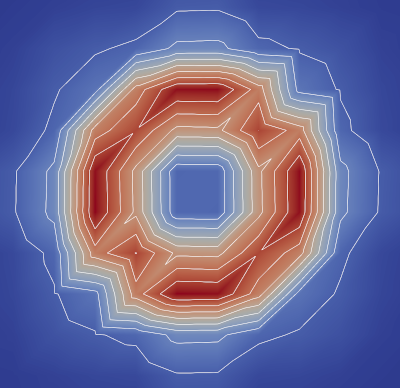}} \quad%
 \subcaptionbox[]{\label{fig:twoPoint_mono}}%
    {\includegraphics[height=3.0cm]{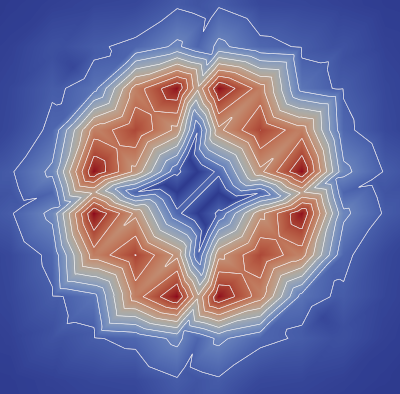}} \quad% 
 \subcaptionbox[]{\label{fig:twoPoint_correct}}%
    {\includegraphics[height=3.0cm]{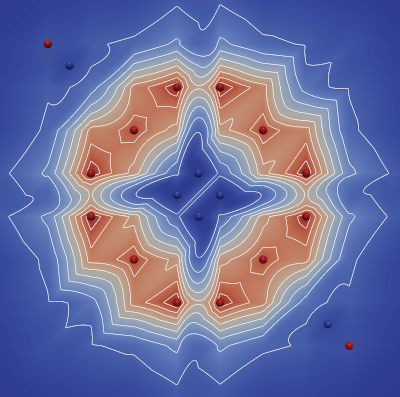}
    \includegraphics[height=3.0cm]{colorMap.png}}\\
 \subcaptionbox[]{\label{fig:twoPoint_cumulative}}%
    {\includegraphics[height=3.5cm]{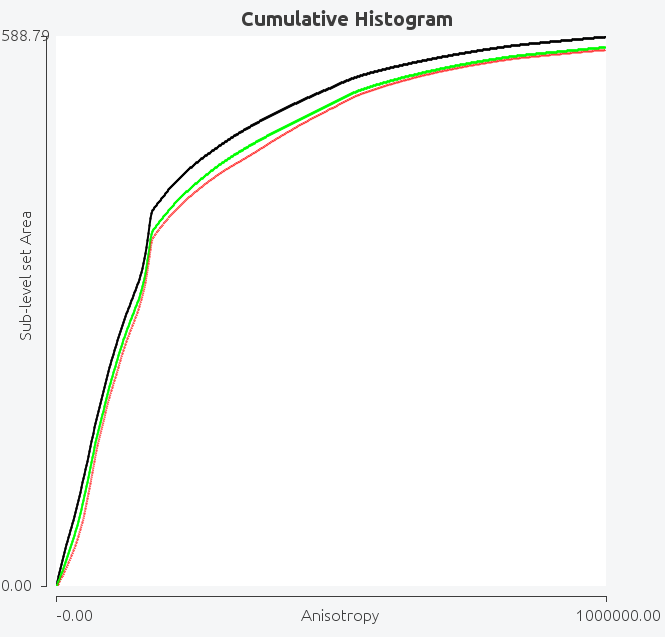}} \quad%
 \subcaptionbox[]{\label{fig:twoPoint_hist}}%
    {\includegraphics[height=3.5cm]{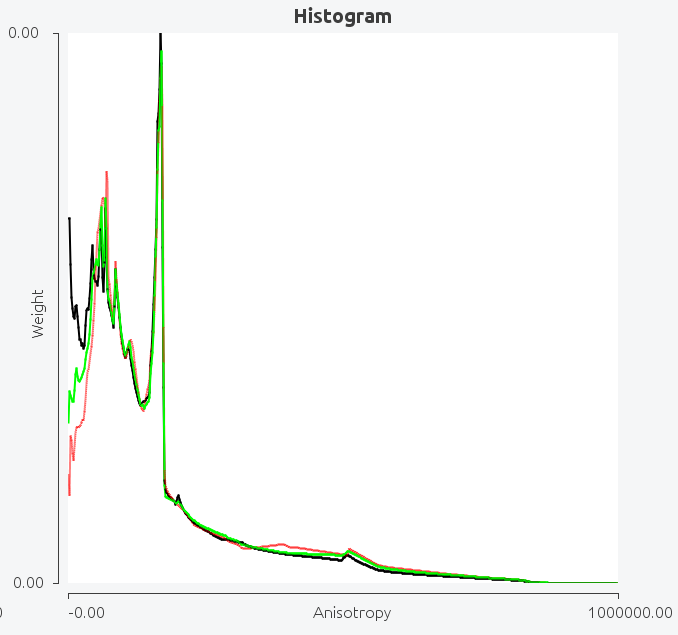}}%
  \caption{Two point load simulation data with focus region around one of the loading points. 
    \subref{fig:twoPoint_orig} The original mesh with anisotropy computed at the mesh vertices and linearly interpolated within the triangles.
    \subref{fig:twoPoint_mono} The subdivided mesh with monotonous triangles. Anisotropy is linearly interpolated within the monotonous triangles.
    \subref{fig:twoPoint_correct} The correct interpolation of anisotropy under linear interpolation of tensor components. The minima (blue) and maxima (red) in the field are highlighted by small spheres. 
    \subref{fig:twoPoint_cumulative} the three cumulative histograms and \protect\subref{fig:twoPoint_hist} the histograms.
    The red curves correspond to (a), the green curves to (b) and the black curve to (c).
    \subref{fig:twoPoint_hist}.
  }
  \label{fig:twoPointLoad_study}
\end{figure}

The difference between the three different interpolations for the anisotropy field gave significantly different results for the synthetic data. In the next step, we want to investigate the impact of these differences on real-world data. At first, we consider a simulation data set from mechanical engineering.
It is a well-known data set, often referred to as two-point load, that represents a stress field in a solid block resulting from the application of two external forces, Figure~\ref{fig:twoPoint_data}. The data is given on a cubic mesh. The anisotropy measure used so far corresponds to the squared von Mises stress, which plays an important role in failure analysis of mechanical components. 
The direction field in one slice is shown in Figure~\ref{fig:two_point_slice}.
In our analysis, we consider a section of this slice, which is shown in Figure~\ref{fig:two_point_slice_select} displaying one eigenvector field. Color represents the anisotropy using a bi-linear interpolation in the original mesh.  In our analysis, we use a triangulated version of the data.

Figure~\ref{fig:twoPointLoad_study} shows the anisotropy fields using the three different interpolations in comparison. In Figure~\ref{fig:twoPoint_orig} one can observe the asymmetry introduced by the triangulation but is similar to~\ref{fig:two_point_slice_select}. The anisotropy fields in~\ref{fig:twoPointLoad_study}\subref{fig:twoPoint_mono}~and~\subref{fig:twoPoint_correct} are both based on the monotonous subdivision and are very similar to each other but differ strongly from~\ref{fig:twoPoint_orig}. The asymmetry due to the triangulation is largely reduced. The minima inside the original triangles in the field capture the locations of the degenerate points of the direction field. 
The corresponding histogram can be seen in Figure~\ref{fig:twoPointLoad_study}\subref{fig:twoPoint_cumulative}~and~\subref{fig:twoPoint_hist}.
As expected they differ strongly for very small values but are rather similar for large values of the anisotropy.

%-------------------------------------------------------
\subsection{Measurement data} 
\begin{figure}[hb]
 \centering
 \includegraphics[width=0.7\linewidth]{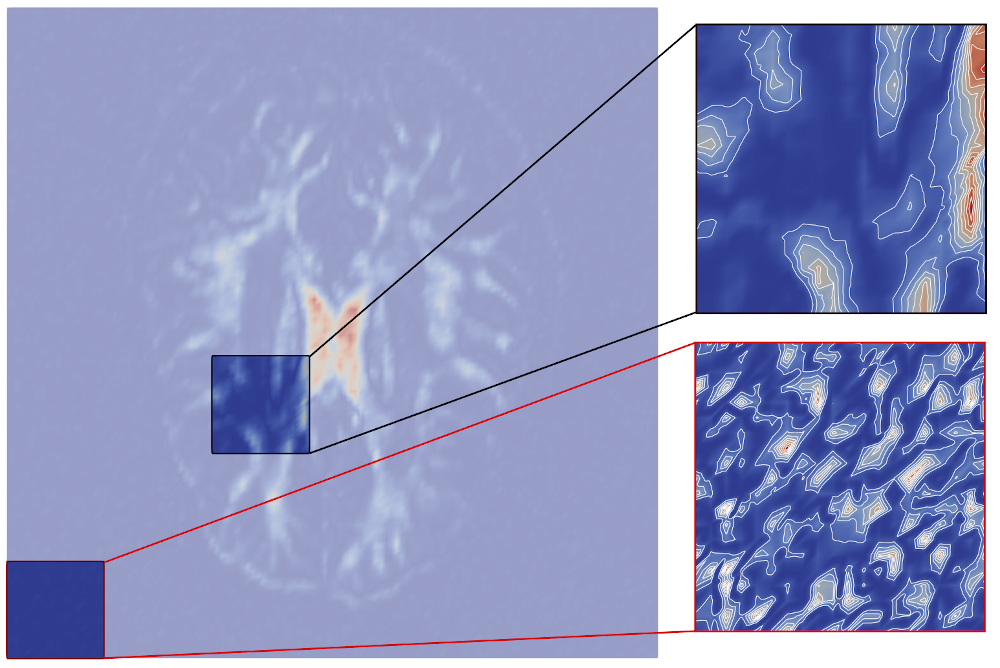}
  \caption{Two segments selected from Diffusion Tensor Imaging data. The 20x20 2D grid shown in red outline is chosen from noisy region of the data while the selection shown in black outline is chosen from the region occupied by the brain. We plot the histograms for these selections in Figures \ref{fig:noise_study} and \ref{fig:dti_data_study} respectively.
  }
  \label{fig:brain_dti}
\end{figure}

As the last example, we examine two sections of a slice of 3D Diffusion Tensor Imaging data. 
Specifically, we compute and compare the anisotropy histograms of noisy regions outside the brain and a region inside the brain. 
The 2D slice from the data along with selected regions is shown in Figure~\ref{fig:brain_dti}. 

\begin{figure}[ht]
 \centering
 \subcaptionbox[]{\label{fig:noise_orig}}%
    {\includegraphics[height=2.6cm]{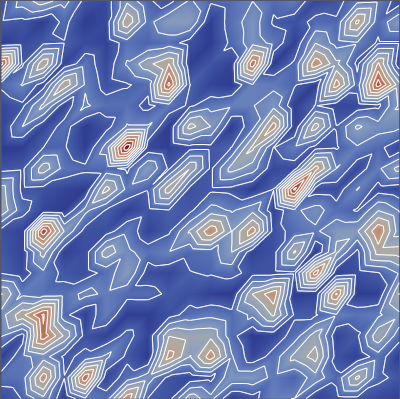}} \quad%
 \subcaptionbox[]{\label{fig:noise_mono}}%
    {\includegraphics[height=2.6cm]{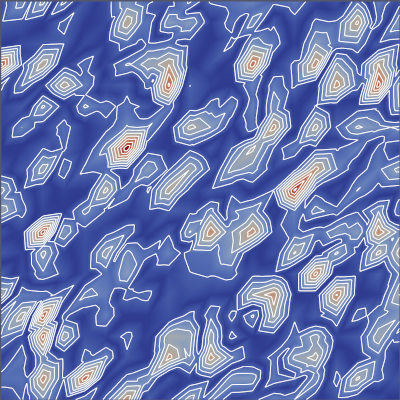}} \quad% 
 \subcaptionbox[]{\label{fig:noise_correct}}%
    {\includegraphics[height=2.6cm]{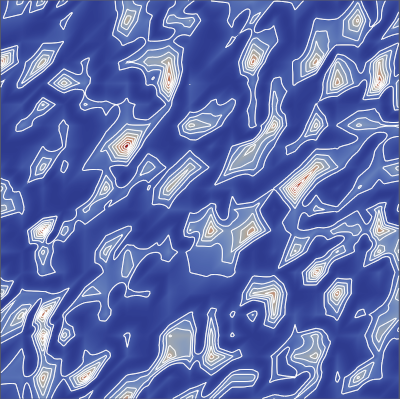}} \\ %
 \subcaptionbox[]{\label{fig:noise_cumulative}}%
    {\includegraphics[height=2.75cm]{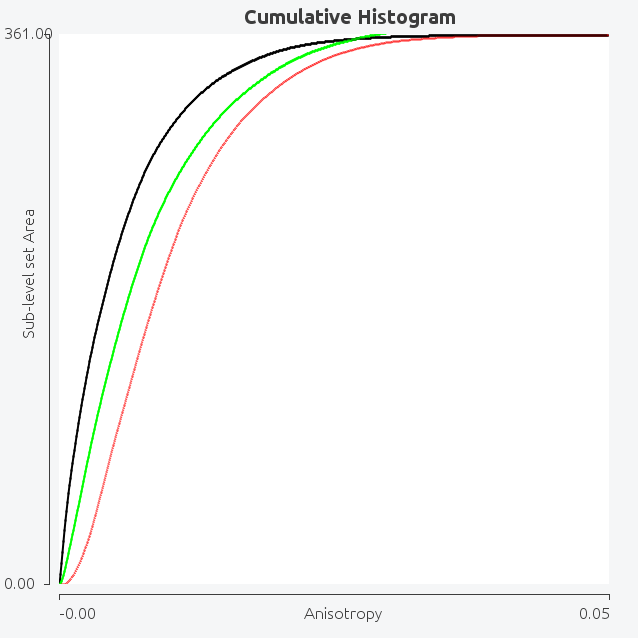}} \quad%
 \subcaptionbox[]{\label{fig:noise_hist}}%
    {\includegraphics[height=2.75cm]{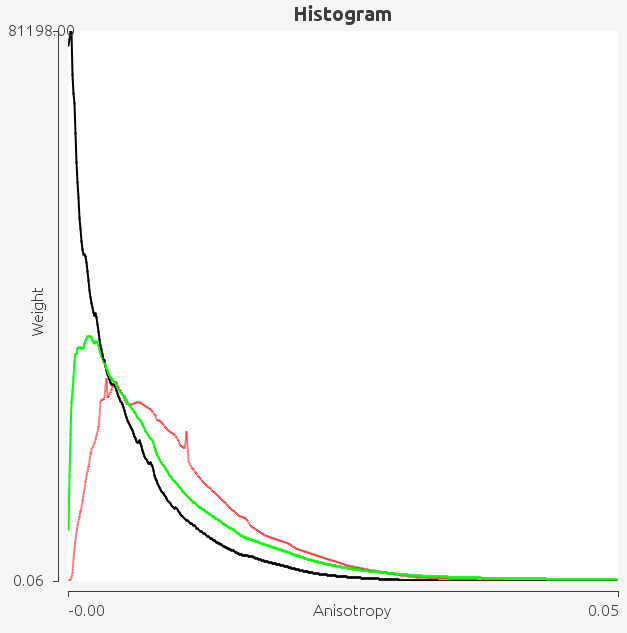}}%
  \caption{Anisotropy fields in a small noisy subset of brain dataset. 
    \subref{fig:noise_orig} Linear interpolation in the original mesh.
    \subref{fig:noise_mono} Linear interpolation within the subdivided mesh with monotonous triangles.
    \subref{fig:noise_correct} The correct interpolation of anisotropy under linear interpolation of tensor components.  
    \subref{fig:noise_cumulative} The comparison of cumulative histograms under these settings. 
    \subref{fig:noise_hist} The comparison of the corresponding histograms with red \subref{fig:noise_orig}, green \subref{fig:noise_mono}, and black \subref{fig:noise_correct}.
  }
  \label{fig:noise_study}
\end{figure}
\begin{figure}[hb]
 \centering
 \subcaptionbox[]{\label{fig:dti_data_orig}}%
    {\includegraphics[height=2.6cm]{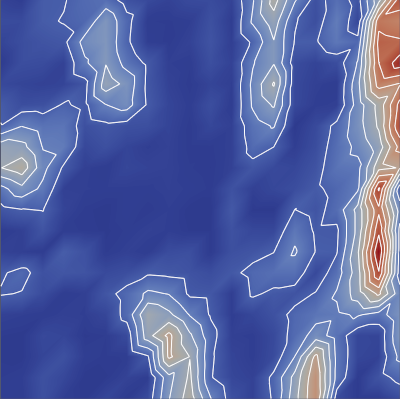}} \quad%
 \subcaptionbox[]{\label{fig:dti_data_mono}}%
    {\includegraphics[height=2.6cm]{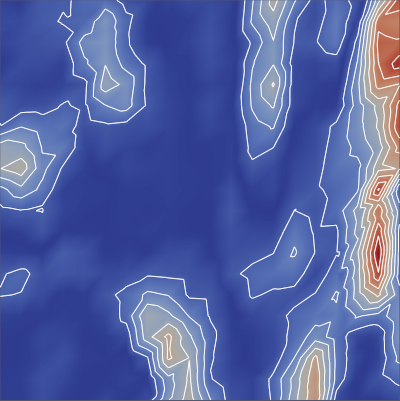}} \quad% 
 \subcaptionbox[]{\label{fig:dti_data_correct}}%
    {\includegraphics[height=2.6cm]{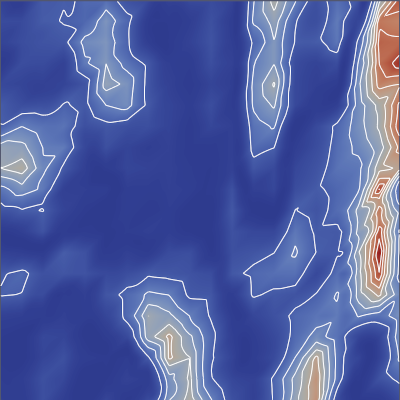}} \\ %
 \subcaptionbox[]{\label{fig:dti_data_cumulative}}%
    {\includegraphics[height=2.75cm]{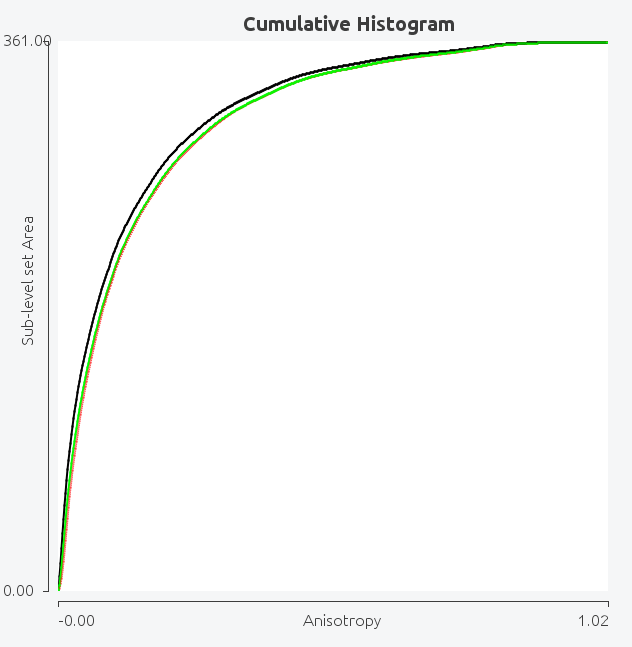}} \quad%
 \subcaptionbox[]{\label{fig:dti_data_hist}}%
    {\includegraphics[height=2.75cm]{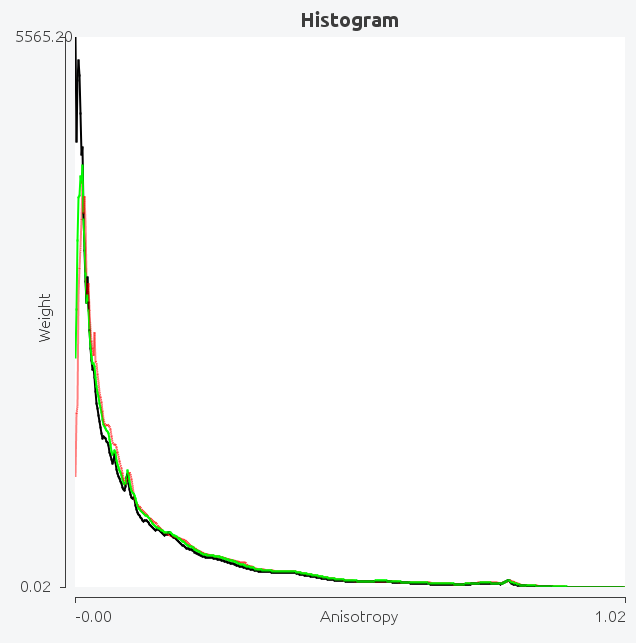}}%
  \caption{A small subset of DTI brain dataset is selected within the brain region. 
    \subref{fig:dti_data_orig} Linear interpolation in the original mesh,
    \subref{fig:dti_data_mono} Linear interpolation in monotonous triangles. 
    \subref{fig:dti_data_correct} The correct interpolation of anisotropy under linear interpolation of tensor components.  
    \subref{fig:dti_data_cumulative} and 
    \subref{fig:dti_data_hist} display the respective cumulative histograms and histograms as
    (a) red, (b) green, (c) black plots.
  }
  \label{fig:dti_data_study}
\end{figure}

Figure~\ref{fig:noise_study} shows the histograms computed for the noisy region. With the interpolation approach [c] for computing histograms, we can observe a high contribution of anisotropy values near zero, which hints at the existence of a lot of degenerate points in the noisy region. This is not captured by the histogram computed with interpolation approach [a]. 
It should be noted that interpolation [b] (green plot) although yields better results than interpolation [a] (red plot), it is still quite different than the correct histogram (black plot).
We did the same experiment for a region that is within the brain and hence contains less noise. The results are shown in Figure~\ref{fig:dti_data_study}. Here we observe, that histograms are very similar whether we use the accurate quadratic function for anisotropy of linearly interpolate it within the triangles.

\clearpage

\section{Conclusions} \label{section:conclusions}
In this paper we explore the behavior of the anisotropy, as an example for a nonlinear derived tensor invariant, when applying a linear component wise interpolation of the tensor field. We demonstrate that a linear interpolation of the invariant itself, the interpolation approach [a], leads to an incorrect topology of contours as well as a bias in the histogram. 
With this analysis we want to emphasize the importance of being consistent with the chosen interpolation in all analysis steps. 
For our analysis we have chosen a linear interpolation of tensor components, which is the most commonly used method in simulations and provides a valid continuous field.  
An independent interpolation of the direction field and the anisotropy violates the preservation of topological invariants and does not result in a valid tensor field.

More specifically we have presented a derivation for the computation of correct contours and  histograms in this setting.
Component wise linear interpolation of tensor components results in a quadratic function for anisotropy. The method is based on a subdivision of the mesh into triangles with monotonous behavior. This subdivision with a linear interpolation of the anisotropy, interpolation approach [b], already results in topologically correct contours. However, the histograms are not accurate and show a bias towards larger anisotropy values. This is especially prominent in regions with many degenerate points. In areas of high anisotropy, interpolation [b] provides a good approximation. The method described in this chapter, interpolation approach [c], can be used to compute correct continuous histogram for anisotropy. 

All derivations in this chapter are given for the anisotropy defined as the difference between the major and minor eigenvalue, of 2D tensor fields. Although not trivial, extension of this work to 3D tensor fields is feasible. An extension to the determinant, which is also quadratic but not elliptic, would be possible in a similar way. A general extension to other non-linear tensor invariants, however might not be possible in a closed form and will require a good approximation schema. Therefore, we plan to explore methods for efficient approximations to the correct distributions with clear error bounds. Also, computing continuous scatterplots to visualize the space of multiple invariants at once is an interesting topic and will be subject of future work.

\bibliographystyle{plain}
\bibliography{references}

\clearpage

%% Appendix
%% ========================================================================= %%
\appendix
\section{Appendix}

\subsection{Detailed analysis of anisotropy and its contours} \label{appendix:anisotropy_analysis}

\subsubsection*{Proof of $H \geq 0$} 
Substituting the equations~\eqref{eq:A} -- \eqref{eq:F} in equation~\eqref{eq:hessianDet}, we can analyse the determinant of Hessian for  anisotropy \sqanis:
\begin{align}
    H =& 4AC - B^2  \nonumber \\ 
       =& 4 \big((e_x-g_x)^2 + 4f_x^2\big) \big((e_y-g_y)^2 + 4f_y^2\big) -  4\big((e_x-g_x)(e_y-g_y) + 4f_xf_y\big)^2 \nonumber \\
       =& 4 \big((e_x-g_x)^2(e_y-g_y)^2 + 4f_x^2(e_y-g_y)^2 + 4f_y^2(e_x-g_x)^2 + 16f_x^2f_y^2 \big) \nonumber \\
        & - 4 \big((e_x-g_x)^2(e_y-g_y)^2 + 8f_xf_y(e_x-g_x)(e_y-g_y) + 16f_x^2f_y^2\big) \nonumber \\
       =& 16 \big( f_x^2(e_y-g_y)^2 + f_y^2(e_x-g_x)^2 - 2f_xf_y(e_x-g_x)(e_y-g_y)\big) \nonumber \\
    H  =& 16 \big( f_x(e_y-g_y) - f_y(e_x-g_x) \big)^2 \geq 0 \label{eq:H_} 
\end{align}
Since $H \geq 0$, we have shown that the contours of \sqanis are never hyperbolic.  

\subsubsection*{Proof of $I = 0$ when $H = 0$}
Let us analyze the case when $H = 0$ to complete the analysis of behaviour of anisotropy in all cases. 
\begin{align}
    H =& 16 \big( f_x(e_y-g_y) - f_y(e_x-g_x) \big)^2 = 0  \nonumber \\ 
or,\quad & f_x(e_y-g_y) - f_y(e_x-g_x) = 0 \nonumber \\
or,\quad & \frac{e_y-g_y}{e_x-g_x} = \frac{f_y}{f_x} \nonumber \\
or,\quad & e_y-g_y = \frac{f_y}{f_x} (e_x-g_x)   \label{eq:rel}
\end{align}
Substituting~\eqref{eq:rel} in~\eqref{eq:C} and using~\eqref{eq:A}:
\begin{align}
    C &= \big((e_y-g_y)^2 + 4f_y^2\big)  \nonumber \\ 
      &= \frac{f_y^2(e_x-g_x)^2}{f_x^2} + 4f_y^2  \nonumber \\
      &= \frac{f_y^2}{f_x^2} \bigg((e_x-g_x)^2 + 4f_x^2\bigg) \nonumber \\
or,\quad C &= \frac{f_y^2}{f_x^2} A \label{eq:CA}
\end{align}
Substituting~\eqref{eq:rel} in~\eqref{eq:E} and using~\eqref{eq:D}:
\begin{align}
    E &= 2\big((e_y-g_y)(e_c-g_c) + 4f_y f_c\big)  \nonumber \\ 
      &= 2\bigg(\frac{f_y(e_x-g_x)(e_c-g_c)}{f_x} + 4f_y f_c\bigg) \nonumber \\
      &= \frac{f_y}{f_x} \cdot 2\big((e_x-g_x)(e_c-g_c) + 4f_xf_c\big) \nonumber \\
 or,\quad E &= \frac{f_y}{f_x} D \label{eq:ED}
\end{align}
From equations~\eqref{eq:CA} and~\eqref{eq:ED}:
\begin{align}
    & \frac{C}{A} = \frac{E^2}{D^2} = \frac{f_y^2}{f_x^2} \label{eq:fracsRel} \\
    or, \quad & AE^2 = CD^2  \label{eq:AECD}
\end{align}
Substituting~\eqref{eq:rel} in~\eqref{eq:B} and using~\eqref{eq:C}:
\begin{align}
    B &= 2\big((e_x-g_x)(e_y-g_y) + 4f_xf_y\big)  \nonumber \\ 
      &= 2\bigg(\frac{f_x(e_y-g_y)^2}{f_y} + 4f_xf_y\bigg) \nonumber \\
      &= 2\frac{f_x}{f_y} \bigg((e_y-g_y)^2 - 4f_y^2\bigg) \nonumber \\
or,\quad B &= 2\frac{f_x}{f_y} C \label{eq:BC}
\end{align}
Multiplying equations~\eqref{eq:ED} and~\eqref{eq:BC}, we obtain:
\begin{equation}
    BE = 2CD \label{eq:BECD}
\end{equation}
Let us evaluate the equation~\ref{eq:I} now:
\begin{align}
    I &= BDE-AE^2-CD^2  &\nonumber \\ 
    I &= BDE-2CD^2      & \text{using} \eqref{eq:AECD} \nonumber \\ 
    I &= D(BE-2CD)      & \nonumber \\
    I &= D(0) = 0       & \text{using} \eqref{eq:BECD} \label{eq:IH_0} 
\end{align}
From equation~\eqref{eq:IH_0}, we conclude that when $H$ is $0$, $I$ is also $0$. This means that the contours of \sqanis~ are never parabolic.

To conclude, based on equations~\eqref{eq:H_} and~\eqref{eq:IH_0}, we deduce that the contours of anisotropy \sqanis~ and hence \anis~ are either ellipses or a set of parallel lines. 

\end{document}